\documentclass[aps,twocolumn,superscriptaddress,longbibliography]{revtex4-2}
\usepackage{array}[=2016-10-06]

\usepackage{latexsym}

\usepackage{cmap}
\usepackage[utf8]{inputenc}
\usepackage{nicefrac}
\usepackage[english]{babel}

\usepackage{amsmath,amssymb}
\usepackage{xcolor}
\usepackage{graphicx}
\usepackage{multirow}
\usepackage{dcolumn}
\usepackage{bm}
\usepackage{braket}
\usepackage{amsthm}
\usepackage[normalem]{ulem}
\usepackage{hyperref}
\usepackage{booktabs}
\setcitestyle{super}

\newtheorem{theorem}{Theorem}
\newtheorem{corollary}{Corollary}

\usepackage{appendix}
\usepackage{chngcntr}
\usepackage{apptools}
\usepackage{multirow}
\usepackage{siunitx}
\usepackage{physics}
\usepackage{tabularx}
\usepackage[version=4]{mhchem}
\usepackage{mathtools}
\usepackage{leftindex}

\usepackage{pifont}

\newcolumntype{L}{>{\raggedright\let\newline\\\arraybackslash\hspace{0pt}}X}
\newcolumntype{C}{>{\centering\let\newline\\\arraybackslash\hspace{0pt}}X}
\newcolumntype{R}{>{\raggedleft\let\newline\\\arraybackslash\hspace{0pt}}X}
\newcolumntype{P}{>{\raggedright\let\newline\\\arraybackslash\hspace{0pt}}p}
\newcolumntype{Y}[1]{>{\hsize=\dimexpr#1\hsize+#1\tabcolsep+\arrayrulewidth\relax\raggedright\let\newline\\\arraybackslash\hspace{0pt}}X}
\AtBeginDocument{\RenewCommandCopy\qty\SI}
\DeclareSIUnit{\Ry}{Ry}
\DeclareSIUnit{\angstrom}{\textup{\AA}}

\newcommand{\siref}[2]{\ref{#1}}

\renewcommand{\thetable}{\arabic{table}}
\newcommand{\citett}[1]{\citeauthor{#1}\cite{#1}}



\begin{document}

\title{Optimizing Quantum Chemistry Simulations with a Hybrid Quantization Scheme}
\author{Calvin Ku}
\affiliation{Hon Hai Research Institute, Taipei, Taiwan}
\affiliation{Department of Mechanical Engineering, City University of Hong Kong, Kowloon, Hong Kong SAR 999077, China}

\author{Yu-Cheng Chen}
\email{kesson.yc.chen@foxconn.com}
\affiliation{Hon Hai Research Institute, Taipei, Taiwan}

\author{Alice Hu}
\email{alicehu@cityu.edu.hk}
\affiliation{Department of Mechanical Engineering, City University of Hong Kong, Kowloon, Hong Kong SAR 999077, China}
\affiliation{Department of Material Science and Engineering, City University of Hong Kong, Kowloon, Hong Kong SAR 999077, China}

\author{Min-Hsiu Hsieh}
\email{min-hsiu.hsieh@foxconn.com}
\affiliation{Hon Hai Research Institute, Taipei, Taiwan}

\begin{abstract}
    Complex quantum simulation workflows are often hindered by incompatible wavefunction representations adopted across different algorithmic frameworks.
    In particular, the mismatch between the first- and second-quantization formalisms prevents algorithms specialized for their respective quantizations from being integrated within a single circuit, thereby forcing practitioners to rely on suboptimal methods simply to maintain a consistent representation.
    To address this challenge, we propose a hybrid quantization scheme that employs a conversion circuit to switch between the two, requiring $\mathcal{O}(N\log N\log M)$ gates for a system of N electrons and M orbitals.
    This capability is critical for constructing complex quantum simulation workflows, allowing us to use the most efficient quantization for each individual step.
    We discuss its applications to bring polynomial improvements in the characterization of ground-state, ab-initio molecular dynamics, and characterization of spectroscopic properties.
    Quantitative estimations of such applications found up to three orders of magnitude fewer ground-state preparations when measuring the 2-reduced density matrix of molecular systems.
\end{abstract}

\maketitle

\noindent\textbf{\large Introduction} \\
\noindent
Performing chemical experiments \emph{in-silico} has been the cornerstone of many scientific discoveries and commercial applications. Research in this direction spans a wide range of areas, including Hamiltonian dynamics, thermochemistry~\cite{cramerEssentialsComputationalChemistry2014}, computational spectroscopy~\cite{kasGreensFunctionsApplied2022,nanniGreensFunctionComputational2023,fomichevFastSimulationsXray2025}, and electron transfer reactions~\cite{wuConstrainedDensityFunctional2006,wuExtractingElectronTransfer2006,kadukConstrainedDensityFunctional2012}.
Most simulations rely on solving eigenstates and applying the time evolution of the Hamiltonian. However, classical algorithms based on the many-body Hamiltonian often have a trade-off in accuracy and efficiency.
The Full Configuration Interaction (FCI)~\cite{szaboModernQuantumChemistry2012} offers high accuracy but is computationally prohibitive, with a cost of $\mathcal{O}(M^{3N})$ for a system of $M$ orbitals and $N$ electrons, restricting its use to small systems. The Coupled Cluster method~\cite{crawfordIntroductionCoupledCluster2000,hirataCoupledclusterSinglesDoubles2004}, commonly with Single and Double excitations (CCSD), provides a practical approximation to FCI, reducing the complexity to $\mathcal{O}(M^3N^3)$. However, even this reduced cost remains impractical for large molecular systems or periodic solids with a large number of orbitals.
Mean-field methods, such as Density Functional Theory (DFT), are widely adopted due to their low computational cost of $\mathcal{O}(M^3)$. However, they often lack the accuracy of more advanced methods and are not well-suited for studying excited states.

\begin{figure*}[!t]
  \centering
  \includegraphics[width=0.95\linewidth]{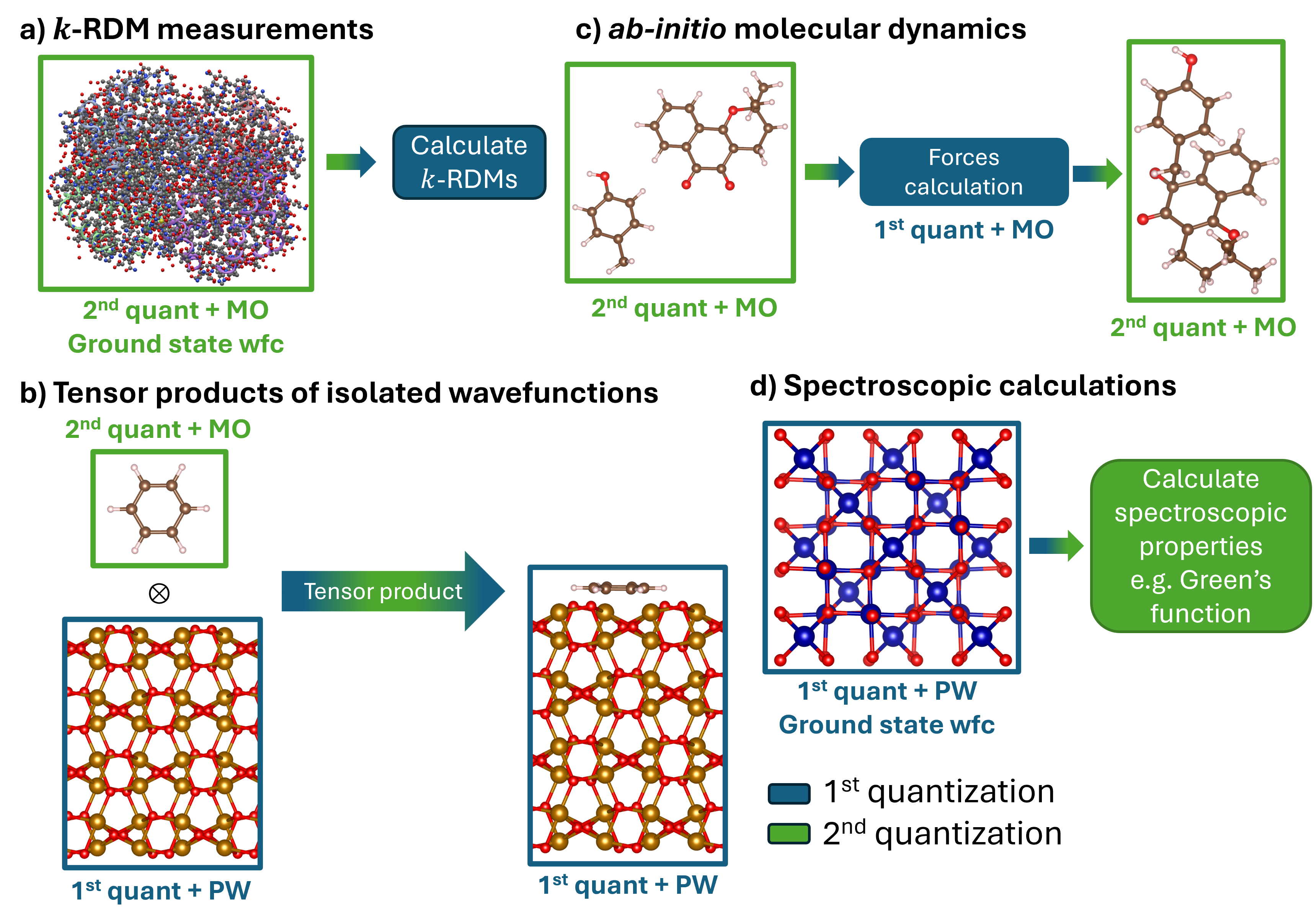}
  \caption{%
    Applications of the hybrid quantization scheme.
    (a) The second-quantized ground-state wavefunction of a Haemoglobin complex is converted to the first-quantization for more efficient $k$-RDM measurements.
    (b) Hybrid quantization is used to perform tensor products of an isolated benzene molecule and an isolated \ce{Fe2O3} surface, eliminating the need to create empty vacuum space that is commonly used to isolate two disentangled systems.
    (c) Efficient AIMD simulations of the product $\beta$-lapachone~\cite{liHybridQuantumComputing2024}.
    (d) The Green's function of a \ce{Co3O4} system was efficiently calculated switching to the first and second quantization.
  }
  \label{fig:illustrations}
\end{figure*}

\begin{table*}[!t]
  \centering
  \caption{%
    Summary of complexity differences between the first-quantized and second-quantized representations for various quantum simulation problems.
    The hybrid quantization framework that we proposed is shown underlined in the Table below, where we show potential workflow paths that result in better Toffoli complexities.
    The variable $N$ represents the number of electrons while $M$ represents the number of orbitals, with $M_{MO}$ and $M_{PW}$ indicating that the cost scaling is specifically for the MO and plane-wave basis, respectively ($M_{MO} \ll M_{PW}$).
    Similarly, $\varepsilon_{\text{RDM}}$, and $\varepsilon_{\text{HAD}}$ are error parameters for the RDM measurements, and Hadamard tests, respectively.
    Finally, the variable $\alpha$ denotes the overlap between the initial state and the ground state.
    For the row ``Ground-state characterization (molecular/bulk system)'', we estimate the cost of measuring all the $k$-RDMs.
    On the other hand, the row ``Ground-state characterization (defect/adsorbed system)'' shows the cost of measuring a small subset of $\mathcal{M}$ observables localised near the defect or adsorbed complex.
  }
  \newcommand{\mrowlength}{3.8cm}
  \begin{tabularx}{\textwidth}{p{\mrowlength}>{\bfseries}r>{\raggedright}p{5cm}L}
    \toprule
    &&Gate cost per application of the ground-state&State preparation complexity\\
    \midrule
    \multirow[t]{3}{\mrowlength}{Ground-state characterization (molecular/bulk system)}
    &1\textsuperscript{st} quant:&$\tilde{\mathcal{O}}(N^{4/3}M_{PW}^{2/3}/\alpha)$~\cite{suFaultTolerantQuantumSimulations2021} &$\mathcal{O}(k^kN^k\log M_{PW}/\varepsilon_{\text{RDM}}^2)$~\cite{babbushQuantumSimulationExact2023} ($N \ll M_{PW}$)\\
    &2\textsuperscript{nd} quant:&$\tilde{\mathcal{O}}(M_{MO}^{2.1}/\alpha)$~\cite{leeEvenMoreEfficient2021} &$\tilde{\mathcal{O}}(M_{MO}^k/\varepsilon_{\text{RDM}})$~\cite{hugginsNearlyOptimalQuantum2022,zhaoFermionicPartialTomography2021} ($N\approx M_{MO}$)\\
    &\underline{Hybrid:}&$\tilde{\mathcal{O}}(M_{MO}^{2.1}/\alpha\ $\cite{leeEvenMoreEfficient2021} $+ N\log N\log M_{MO})$&$\mathcal{O}(k^kN^k\log M_{MO}/\varepsilon_{\text{RDM}}^2)$~\cite{babbushQuantumSimulationExact2023} ($N\ll M_{MO}$)\\
    \cmidrule(lr){1-4}
    \multirow[t]{3}{\mrowlength}{Ground-state characterization (defect/adsorbed system)}
    &1\textsuperscript{st} quant:&$\tilde{\mathcal{O}}(N^{4/3}M_{PW}^{2/3}/\alpha)$~\cite{suFaultTolerantQuantumSimulations2021} &$\mathcal{O}(k^kN^k\log M_{PW}/\varepsilon_{\text{RDM}}^2)$~\cite{babbushQuantumSimulationExact2023} \\
    &2\textsuperscript{nd} quant:&$\tilde{\mathcal{O}}(M_{MO}^{2.1}/\alpha)$~\cite{leeEvenMoreEfficient2021} &$\tilde{\mathcal{O}}(\sqrt{\mathcal{M}}/\varepsilon_{\text{RDM}})$~\cite{hugginsNearlyOptimalQuantum2022,zhaoFermionicPartialTomography2021} \\
    &\underline{Hybrid:}&$\tilde{\mathcal{O}}(N^{4/3}M_{PW}^{2/3}/\alpha + N\mathcal{M}M_{PW})$~\cite{suFaultTolerantQuantumSimulations2021}
    &$\tilde{\mathcal{O}}(\sqrt{\mathcal{M}}/\varepsilon_{\text{RDM}})$~\cite{hugginsNearlyOptimalQuantum2022,zhaoFermionicPartialTomography2021} (PW $\rightarrow$ MO)\\
    \cmidrule(lr){1-4}
    \multirow[t]{3}{\mrowlength}{Born-Oppenheimer~\cite{bornZurQuantentheorieMolekeln1927} AIMD simulation}
    &2\textsuperscript{nd} quant:&$\tilde{\mathcal{O}}(M_{MO}^{2.1}/\alpha)$&$\tilde{\mathcal{O}}(M_{MO}^2/\varepsilon_{\text{RDM}})$\\
    &\underline{Hybrid}:&$\tilde{\mathcal{O}}(M_{MO}^{2.1}/\alpha + N\log M_{MO}\log\log M_{MO})$&$\tilde{\mathcal{O}}(N^2\log M_{MO}/\varepsilon_{\text{RDM}}^2)$\\
    \cmidrule(lr){1-4}
    \multirow[t]{3}{\mrowlength}{Spectroscopic characterization (Eqn.~\ref{eqn:greens_2})}
    &1\textsuperscript{st} quant:&Not applicable&Not applicable\\
    &2\textsuperscript{nd} quant:&$\tilde{\mathcal{O}}(M_{PW}^{7/3})$&$\mathcal{O}(1/\varepsilon_{\text{HAD}}^2)$~\cite{linLectureNotesQuantum2022} (PW dual basis~\cite{childsQuantumAlgorithmSystems2017})\\
    &\underline{Hybrid:}&$\tilde{\mathcal{O}}(NM_{PW}^{2/3})$&$\mathcal{O}(1/\varepsilon_{\text{HAD}}^2)$~\cite{linLectureNotesQuantum2022} \\
    \bottomrule
  \end{tabularx}
  \label{tab:intro_table}
\end{table*}

Recent advances in quantum computing technologies have sparked interest in developing quantum algorithms to tackle quantum chemistry challenges.
A distinctive feature of quantum computers is the ability of $M$ qubits to be in a superposition of $2^M$ distinct values.
This capability enables efficient encoding of both first- and second-quantized many-body wavefunctions.
Using this superposition of qubits, quantum algorithms like Quantum Phase Estimation (QPE)~\cite{nielsenQuantumComputationQuantum2012,georgesQuantumSimulationsChemistry2025,babbushQuantumSimulationChemistry2019,suFaultTolerantQuantumSimulations2021,babbushEncodingElectronicSpectra2018,kimFaulttolerantResourceEstimate2022} can operate simultaneously on these states, potentially surpassing classical algorithms in scaling efficiency to solve equilibrium properties for chemical systems.
Beyond efficient wavefunction representation, quantum advantage in chemical simulations also stems from the efficient implementation of the Hamiltonian.
This is typically achieved via either trotterization or qubitization, both of which are extensive studied in the literature.
For trotterization, significant research efforts are dedicated to obtaining tight bounds of the trotter error to minimize the required step size to achieve a desired precision.
While errors bounds can be derived from the commutators of the Hamiltonian~\cite{childsTheoryTrotterError2021}, these are often challenging to calculate for Hamiltonians in the molecular orbital (MO) bases~\cite{guntherPhaseEstimationPartially2025}.
Conversely, more structured bases, such as the plane-wave and its dual basis, yield a more structured Hamiltonian.
Hence, despite requiring a larger number of orbitals for comparable precision, tight bounds of the trotter error can be calculated~\cite{babbushLowDepthQuantumSimulation2018,suNearlyTightTrotterization2021}, leading to an improved asymptotic scaling compared to the MO basis.
Furthermore, studies have also investigated Trotter error analysis for Coulomb operators without spatial discretization~\cite{fangTrotterErrorManybody2025}, governing the trotter error bounds when the spatial discretization tends to infinity.
For the case of qubitization, the majority of the research focuses on rank-reduction techniques~\cite{berryImprovedTechniquesPreparing2018,vonburgQuantumComputingEnhanced2021,leeEvenMoreEfficient2021}. These techniques employs method similar to singular value decompositions (SVD) in tensor networks to reduce the gate costs associated with implementing the qubitization operators.
Similarly with the trotterization case, specialized circuits for the plane-wave basis can be constructed with significantly lower asymptotic scaling with respect to the number of orbitals~\cite{babbushQuantumSimulationChemistry2019,suFaultTolerantQuantumSimulations2021}.

For preparing the eigenstate, the most efficient algorithm for the plane-wave basis-set is done in the first quantization with a Toffoli complexity of $\tilde{\mathcal{O}}(N^{8/3}M^{1/3}+N^{4/3}M^{2/3})$~\cite{babbushQuantumSimulationChemistry2019,suFaultTolerantQuantumSimulations2021}, where we use the notation $\tilde{\mathcal{O}}$ to ignore poly-logarithmic factors.
In contrast, the second-quantized representation provides a more efficient algorithm for the molecular orbital (MO) basis, with a Toffoli complexity of $\mathcal{O}(M^{2.1})$~\cite{leeEvenMoreEfficient2021}.
Efficient algorithms for measuring chemical properties, such as the $k$-reduced density matrices~\cite{babbushQuantumSimulationExact2023,hugginsNearlyOptimalQuantum2022,huangPredictingManyProperties2020,lowClassicalShadowsFermions2024,ogormanFermionicTomographyLearning2022,zhaoFermionicPartialTomography2021} (RDMs), have also been developed for both quantizations.
In the first quantization, all the $k$-RDMs can be estimated with $\mathcal{O}(k^kN^k\log M/\varepsilon^2)$ measurements using classical shadows~\cite{babbushQuantumSimulationExact2023}, given a specified $k$ and an allowable error of $\varepsilon$. In the second quantization, amplitude estimation~\cite{hugginsNearlyOptimalQuantum2022} achieves the same with $\tilde{\mathcal{O}}(M^k/\varepsilon)$ measurements.

Beyond solving equilibrium properties, the simulation of material and molecular non-equilibrium properties, such as the Green's functions and \emph{ab-initio} molecular dynamics (AIMD), requires additional operations beyond eigenstate preparation and chemical properties measurements.
Green's function calculations~\cite{fomichevFastSimulationsXray2025,harrowQuantumAlgorithmLinear2009,tongFastInversionPreconditioned2021,caiQuantumComputationMolecular2020,chenVariationalQuantumEigensolver2021,runggerDynamicalMeanField2020} are crucial for studying the spectroscopic properties~\cite{nanniGreensFunctionComputational2023}, which can be experimentally validated using techniques such as photoemission spectroscopy~\cite{chenVariationalQuantumEigensolver2021,fomichevFastSimulationsXray2025}.
These calculations involve ground-state preparation, electron non-conserving operators, Hamiltonian inversion, and Hadamard tests.
Similarly, AIMD simulations iterate through the electronic ground state preparation, ionic forces calculation, ionic simulation, and electronic Hamiltonian update steps, offering a first-principles approach to phenomena like hydrogen bonding, unlike classical semi-empirical methods~\cite{grimmeSemiempiricalGGAtypeDensity2006,grimmeConsistentAccurateInitio2010}.
While VQE-based AIMD~\cite{laiAccurateEfficientCalculations2023,tillyVariationalQuantumEigensolver2022} for noisy intermediate-scale quantum devices has been explored, fault-tolerant quantum-based AIMD, even with pseudo-ion approximation~\cite{jornadaComprehensiveFrameworkSimulate2025}, allows researchers to explore larger and more complex systems, such as enzymatic reactions, nanoparticle formation~\cite{gaoUnderstandingInterfacialNanoparticle2022}, and atmospheric degradation of volatile organic compounds~\cite{vondomarosMolecularDynamicsSimulations2025}.

For these types of calculations, however, each step of the simulation workflow must be scrutinized for potential bottlenecks when selecting an appropriate quantization.
To tackle these inefficiencies, we propose conversion circuits that enable seamless transitions between quantization methods, shown in \autoref{thm:conversion}.
This approach allows one to adopt the most efficient quantization for each simulation step, rather than adhering to a single method throughout the entire workflow.
This hybrid scheme is therefore indispensable for achieving optimal scaling of such complex workflows by providing a robust and efficient method to seamlessly convert between the two quantizations.
To increase the variety of potential applications for our hybrid quantization framework, we also introduced \autoref{coro:basis_transformation}, which allows efficient transformations between the plane-wave and the MO basis in the first quantization.
We then showcase applications of our framework as illustrated in \autoref{fig:illustrations}, including applications in ground-state materials characterization for both bulk and defect systems, improving the cost of AIMD simulations, and for the characterization of spectroscopic and electron ionization and attachment properties of materials. These applications mostly show polynomial improvements over the previous solutions as shown in \autoref{tab:intro_table}.

\medskip
\noindent\textbf{\large Results}\\
\noindent\textbf{Hybrid Quantization Framework}\\
The many-body wavefunction of fermionic systems can be expressed in the first or the second quantization, both serving distinct purposes.
The first quantization straightforwardly expresses the many-body wavefunction as a linear superposition of Hartree products.
These Hartree products are tensor products of the respective single-particle wavefunction.
It only expresses wavefunctions with the same number of particles, simplifying the construction of operators.
This led to efficient Hamiltonian simulations for certain bases, such as plane waves.
However, as arbitrary superpositions of Hartree products may not necessarily obey the Pauli exclusion principle, wavefunctions expressed in the first quantization must be explicitly antisymmetrized.
Conversely, the second quantization aims for an antisymmetrized basis for the many-body wavefunction.
Hence it expresses the wavefunction as a linear superposition of Slater determinants.
This way, any arbitrary wavefunction automatically obeys the Pauli exclusion principle.
Moreover, the second quantization also allows wavefunctions of differing occupation numbers to be in a superposition.
This has the trade-off of requiring the operators to maintain parity of the wavefunction, which complicates second-quantized operator implementations.
To apply these settings, different fermion-to-qubit encodings~\cite{jordanUeberPaulischeAequivalenzverbot1928,bravyiFermionicQuantumComputation2002,seeleyBravyiKitaevTransformationQuantum2012,bravyiTaperingQubitsSimulate2017,kirbyQuantumSimulationSecondquantized2021,kirbySecondQuantizedFermionicOperators2022,sheeQubitefficientEncodingScheme2022,steudtnerFermiontoqubitMappingsVarying2018,chengOptimalNumberconservedLinear2025,babbushExponentiallyMorePrecise2018,carolanSuccinctFermionData2024} have been proposed.

For accurate calculations of the electronic structure, orbitals are required to vastly outnumber electrons as $N\ll M$.
The optimal encoding, where both qubit and gate complexity are logarithmically dependent on $M$, can be realized using variants of the binary addressing code~\cite{steudtnerFermiontoqubitMappingsVarying2018} for both the first~\cite{babbushQuantumSimulationChemistry2019,suFaultTolerantQuantumSimulations2021} and second~\cite{carolanSuccinctFermionData2024} quantizations.
The first optimal variant is the ``first-quantized encoding'', used for first-quantized Hamiltonian simulations that are specified in the plane-wave basis~\cite{kassalPolynomialtimeQuantumAlgorithm2008,babbushQuantumSimulationChemistry2019,suFaultTolerantQuantumSimulations2021}. It uses $N$ registers, each comprising $\mathcal{O}(\log M)$ qubits, to store the binary representation of the orbital indices, resulting in a qubit scaling of $\mathcal{O}(N\log M)$.
State-of-the-art techniques in the first quantization also leverage this code with $\mathcal{O}(\log M)$ gate operations (e.g., binary arithmetic and comparison) per register, totalling a gate cost of $\mathcal{O}(N\log M)$. A unique advantage of the first quantization is its efficiency in the plane-wave basis, enabling large material simulations. Another optimal variant is the ``sorted-list encoding''~\cite{carolanSuccinctFermionData2024}, used for second-quantized Hamiltonian simulations with Majorana operators.
This mitigates the inefficiency of the binary addressing code, where the decomposition of fermionic operators to Pauli operators is prohibitively expensive due to its non-linear nature~\cite{steudtnerFermiontoqubitMappingsVarying2018}. An ascending-order constraint ensures Slater determinants remain order-independent, and fermionic operators now match the $\mathcal{O}(N\log M)$ gate cost of first quantization. A unique advantage of the second quantization is the encoding of ``unoccupied'' orbitals, enabling representation of Slater determinants with different electron numbers and supporting electron non-conserving operators. Further details on both quantizations can be found in \siref{sec:prev_encoding}{I}.

The two encodings described above serves different purposes in the two quantizations, but the similar binary addressing encoding bridges the two quantization. This insight enables efficient conversion between the two quantizations realized via \autoref{thm:conversion}, allowing a hybrid approach that switches between quantizations mid-simulation to exploit the strengths of each quantizations while avoiding their limitations, showcasing a drastic improvement on resource requirement in material and molecular simulation.

\begin{theorem}
  \label{thm:conversion}
  Conversion between the first-quantized encoding and the sorted-list encoding can be performed with a gate cost of $\mathcal{O}(N\log N\log M)$ with a qubit complexity of $\mathcal{O}(N\log M)$.
\end{theorem}
\begin{proof}
  The proof can be found in the Methods section.
\end{proof}

It is important to note that two distinct circuits are required for the two conversion directions, as neither circuits can be straightforwardly reversed to perform the opposite conversion.
Specifically, the conversion from the second-quantized sorted-list encoding to the first-quantized encoding requires postselection of ancilla qubits~\cite{berryImprovedTechniquesPreparing2018}, which prevents a reversible operation.
Conversely, the conversion from the first-quantized to the sorted-list encoding outputs ancilla qubits at an arbitrary state.
Fortunately, the specific state of those ancilla qubits is unentangled from the data qubits and can be safely discarded, as shown in the proof of \autoref{thm:conversion}.
However, as we lack access to those specific ancilla states when implemented in the reverse direction, this conversion circuit is also not reversible.

In order to leverage the hybrid quantization framework to its full potential, we also introduced basis transformation in the first quantization.
Consider a single-particle basis transformation matrix $u$ of dimension $M_1\times M_2$.
The application of this basis transformation matrix facilitates the conversion between different bases, such as from the MO basis to the plane-wave basis.
This transformation can be expressed in first and second quantization as
\begin{equation*}
  \psi_i^{(1)} = \sum_j u_{ij} \psi_j^{(2)} \quad\text{and}\quad (a^{(1)}_i)^\dagger = \sum_j u_{ij}(a^{(2)}_j)^\dagger,
\end{equation*}
respectively.
Using the Thouless theorem~\cite{thoulessStabilityConditionsNuclear1960}, we would first express the global transformation matrix $\mathbb{U}(u)$ for the Slater determinant as
\begin{equation}
  \mathbb{U}(u) = \exp\left(\sum_{ij}[\log u_{ij}](a_i^\dagger a_j - a_j^\dagger a_i)\right).\label{eqn:thouless}
\end{equation}
This can then be decomposed into $\mathcal{O}(M_1M_2)$ Pauli rotations~\cite{kivlichanQuantumSimulationElectronic2018}, resulting in a total gate complexity of $\mathcal{\tilde{O}}(M_1M_2)$ for the Jordan-Wigner encoding, and $\mathcal{O}(M_1M_2N\log M)$ for the sorted-list encoding.
However, when the single-particle basis-transformation matrix $u$ is non-unitary, the operation becomes non-trivial.
Following the work of \citett{zhuQuantumCircuitNonUnitary2025}, the key insight is that $\mathbb{U}$ preserves the composition of the basis transformation matrices:
\begin{equation*}
  u = u_1 u_2  \implies \mathbb{U}(u) = \mathbb{U}(u_1) \mathbb{U}(u_2).
\end{equation*}
This can then be done to the singular value decomposition (SVD) of $u$
\begin{equation*}
  u = LDR \implies \mathbb{U}(u) = \mathbb{U}(L)\mathbb{D}\mathbb{U}(R),
\end{equation*}
where $L$ and $R$ are $M_1\times M_1$ and $M_2\times M_2$ unitary matrices, respectively, and $D$ is an $M_1\times M_2$ diagonal matrix.
Since $L$ and $R$ are unitary, $\mathbb{U}(L)$ and $\mathbb{U}(R)$ can be implemented via the Thouless theorem of \autoref{eqn:thouless}.
On the other hand, the conversion of the $D$ case is shown in \citett{zhuQuantumCircuitNonUnitary2025}.
As a result, this transformation can be performed with a cost of $\mathcal{O}(\max(M_1,M_2)^2)$~\cite{zhuQuantumCircuitNonUnitary2025} for the Jordan-Wigner encoding.
In contrast, basis transformations in the first quantization is detailed in \autoref{coro:basis_transformation}, with more straightforward extensions to the non-unitary case~\cite{zhuQuantumCircuitNonUnitary2025}.

\begin{corollary}
  \label{coro:basis_transformation}
  Basis transformations in the first-quantized encoding can be performed by applying the single-particle basis transformation matrix once for each of the $N$ registers.
  Those registers contains $\lceil\log_2 M\rceil$ qubits each, which stores the binary representation of the orbital index occupied by each of the $N$ electronic degrees of freedom.
  As a result, implementing a $M_1\times M_2$ basis transformation matrix would, in the worst-case, involve single-qubit gates and $\mathcal{O}(NM_1M_2)$ CNOT gates~\cite{itenQuantumCircuitsIsometries2016}.
\end{corollary}
\begin{proof}
  The proof can be found in the Methods section.
\end{proof}

While the complexity in \autoref{coro:basis_transformation} represents a worse-case scenario, practical implementations are often less costly.
For example, converting between the plane-wave and plane-wave dual basis in second quantization uses the Fermionic Fast Fourier Transform (FFFT) circuit~\cite{babbushLowDepthQuantumSimulation2018}, achieving a gate cost of $\mathcal{O}(M^2)$.
In first quantization, the same transformation leverages the Quantum Fourier Transform (QFT)~\cite{coppersmithApproximateFourierTransform2002,halesImprovedQuantumFourier2000} applied to each of the $N$ registers spanning $\mathcal{O}(\log M)$ qubits, resulting in a gate cost of $\mathcal{O}(N\log M\log\log M)$.

Alternatively, when the circuit implementation of $u$ is costly, we can leverage the technique discussed in \citett{babbushQuantumSimulationExact2023} to achieve a gate cost of $\mathcal{O}(NM_1\log M_1 + NM_2\log M_2 + M_1M_2)$.
This is achieved by converting the first-quantized encoding to the Jordan-Wigner encoding.
This conversion proceeds by first applying \autoref{thm:conversion} to temporarily convert to the sorted-list encoding, and then utilizing the circuits detailed in Supplementary Note 7 of~\citett{babbushQuantumSimulationExact2023} to obtain the Jordan-Wigner encoding.
The basis transformation itself is performed in the Jordan-Wigner encoding with a gate cost of $\mathcal{O}(M_1M_2)$, followed by a conversion back to the first-quantized encoding.
A notable limitation of this method is the $\mathcal{O}(\max(M_1,M_2))$ ancilla requirement due to the use of the Jordan-Wigner encoding, rendering it impractical for large basis-sets such as the plane-wave basis.

\medskip
\noindent\textbf{\large Applications}\\
In this section, we explore the hybrid quantization framework, an innovative approach that merges the strengths of first and second quantization to tackle complex quantum problems with greater efficiency and flexibility.
This framework is particularly useful when traditional methods fall short, offering a powerful tool for advanced simulations.
We will dive into three key scenarios where it proves invaluable: ground-state materials characterization, Born-Oppenheimer AIMD simulation, and characterization of spectroscopic and electron ionization and attachment properties.
A cartoon depiction of these applications is illustrated in \autoref{fig:illustrations}.

\medskip
\noindent\textbf{Ground-state characterization (molecular/bulk system).}
Characterization of materials using quantum computers involves two major steps: Hamiltonian simulation to prepare the ground state, and the measurement of observables with respect to the ground state.
Examples of such observables are ionic forces used in molecular dynamics simulation, and the Hessian matrix of the Hamiltonian used in obtaining vibrational frequencies for computational thermochemistry.
As such observables are typically expressed in terms of the second-quantized creation and annihilation operators, their expectation values can be calculated by measuring the reduced density matrices (RDMs) of the ground state wavefunctions
\begin{equation*}
    O = \sum_{pq} o_{pq} a_p^\dagger a_q\implies \ev{\hat{O}} = \sum_{pq} o_{pq} \ev{a_p^\dagger a_q}.
\end{equation*}
While the observable above only contains one-electron terms, this can be generalized to the expectation value of $k$-electron terms called $k$-RDMs
\begin{equation*}
  \leftindex^{k} D^{p_1,\dots,p_k}_{q_1,\dots,q_k} = \ev{a_{p_1}^\dagger \cdots a_{p_k}^\dagger a_{q_k}\cdots a_{q_1}}.
\end{equation*}
Many algorithms have been developed to estimate the $k$-RDMs, including sampling~\cite{jenaOptimization2022,yenMeasuringAllCompatible2020,verteletskyiMeasurementOptimizationVariational2020}, amplitude estimation~\cite{brassardQuantumAmplitudeAmplification2002,ogormanFermionicTomographyLearning2022}, gradient calculation~\cite{hugginsNearlyOptimalQuantum2022}, shadow tomography~\cite{aaronsonShadowTomographyQuantum2020,huangInformationTheoreticBoundsQuantum2021}, and classical shadows~\cite{huangPredictingManyProperties2020,zhaoFermionicPartialTomography2021,babbushQuantumSimulationExact2023}.
In general, when $N$ is not much smaller than $M$, the gradient-based method~\cite{hugginsNearlyOptimalQuantum2022} is optimal with $\tilde{\mathcal{O}}(M^k/\varepsilon)$ applications of the ground-state circuit.

\begin{figure}[h]
  \centering
  \includegraphics[width=\linewidth]{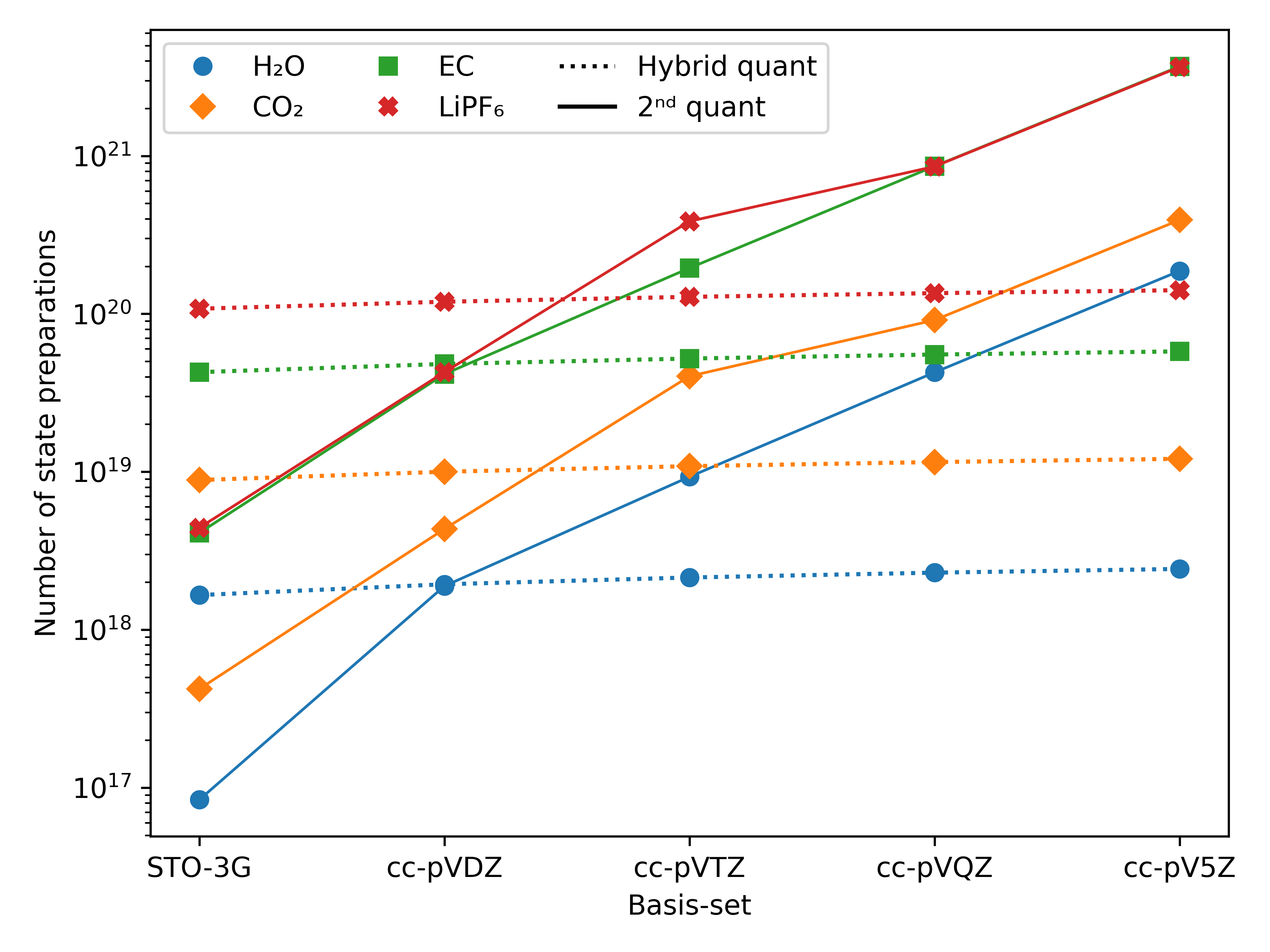}
  \caption{%
    A plot of the number of state-preparations required to obtain the $2$-RDM for several molecular systems.
    This is calculated with 99\% success probability such that each of the $2$-RDM has error of less than $10^{-5}$ Ha, and includes \ce{H2O}, \ce{CO2}, ethylene carbonate \ce{(CH2O)2CO} (EC), and \ce{LiPF6}.
    The solid lines labelled ``2\textsuperscript{nd} quant'' performs ground-state preparation and the gradient-based RDM measurement algorithm, both in the second quantization.
    The dashed lines labelled ``Hybrid quant'' perform ground-state preparation in the second quantization, then uses the first-quantized classical shadows RDM measurement algorithm, with \autoref{thm:conversion} in between to convert between the two quantizations.
    Details on the calculation is discussed in \siref{sec:rdm_formula}{II}.
  }
  \label{fig:rdm_plot}
\end{figure}

In the $N\ll M$ regime, $k$-RDM measurements are more efficient using classical shadows.
It can measure the set of all $k$-RDMs in the first quantization up to an error of $\varepsilon$ by measuring $\mathcal{O}(k^kN^k\log M/\varepsilon^2)$ preparations of the ground-state~\cite{babbushQuantumSimulationExact2023}.
As a result, the second-quantized sorted-list encoding combined with the first-quantized classical shadows, using \autoref{thm:conversion}, has the advantage over the purely second-quantized gradient-based measurement algorithm, as shown in the first row of ~\autoref{tab:intro_table}.
In total, this strategy can achieve a total best-case Toffoli cost scaling of $\mathcal{O}(M^{2.1}k^kN^k\log M/\varepsilon^2\alpha)$, where $\varepsilon$ represents the precision parameter, and $\alpha = |\langle\Psi_{\text{init}}|\Psi_g\rangle|$ represents the overlap between the initial state and the ground state.
For comparison, classical calculations of obtaining the ground-state and then measuring the $2$-RDM costs $\mathcal{O}(M^N)$ for the FCI calculation~\cite{coeEfficientComputationTwoElectron2022} and $\mathcal{O}(N^4M^4)$ for the CCSD calculation~\cite{gaussCoupledclusterOpenshellAnalytic1991,gaussAnalyticEvaluationEnergy1991}.

\begin{figure}[t]
  \begin{center}
    \includegraphics[width=\linewidth]{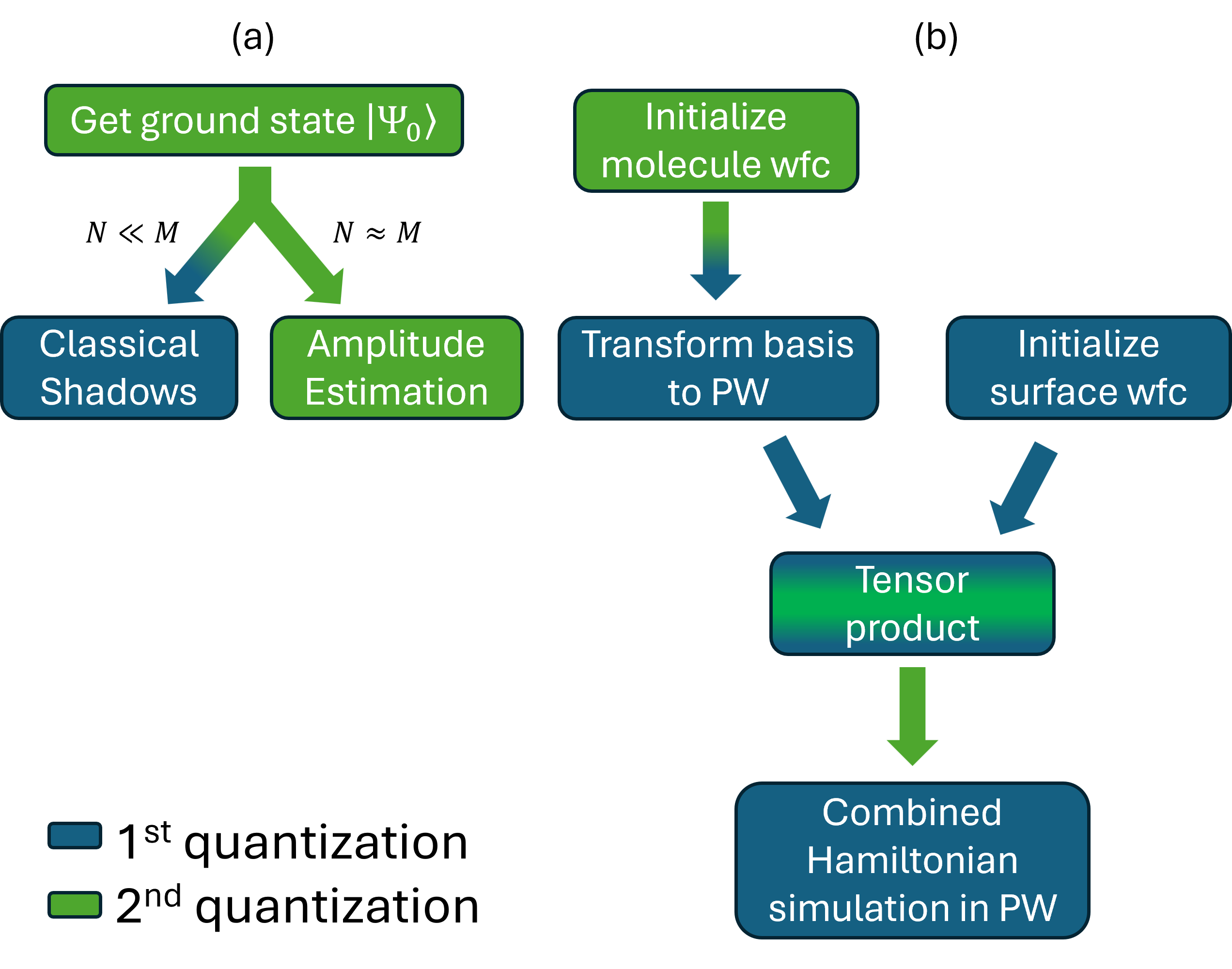}
  \end{center}
  \caption{%
    Flowcharts to illustrate the workflow of materials characterization.
    (a) Workflow to measure reduced density matrices.
    (b) Workflow to perform a tensor product between the wavefunction of a molecule in the MO basis and the wavefunction of a surface in the plane-wave basis.
  }
  \label{fig:material_characterization}
\end{figure}

This is demonstrated numerically in \autoref{fig:rdm_plot}, where we plotted the number of state preparations required to obtain the $2$-RDM for several molecular system.
For smaller basis-sets, specifically STO-3G and cc-pVDZ, the second-quantized measurement algorithm requires fewer applications of the ground-state preparation.
However, larger basis-sets favour the first-quantized classical shadows algorithm in terms of the required number of ground-state preparations, necessitating the use of \autoref{thm:conversion} to convert the second-quantized ground-state to the first quantization.
In the extreme case of the cc-pV5Z basis, this corresponds to around 20-80 times fewer ground-state for the molecules shown in \autoref{fig:rdm_plot}.
As a result, conversion to the first-quantized encoding enables more efficient measurements via classical shadows, as shown in \autoref{fig:material_characterization}(a).
While similar asymptotic scaling can be achieved with another second-quantized classical shadows algorithm~\cite{lowClassicalShadowsFermions2024}, implementing this for the sorted-list encoding would also involve the hybrid quantization scheme.
This is elaborated further in \siref{sec:classical_shadows}{III}.

\medskip
\noindent\textbf{Ground-state characterization (defect/adsorbed system).}
But this does not imply that every calculation in the $N\ll M$ regime should employ the first-quantized classical shadow algorithm.
The second-quantized gradient-based algorithm~\cite{hugginsNearlyOptimalQuantum2022} exhibits better scaling with respect to $\varepsilon$ and allows an arbitrary set of observables, where $\mathcal{M}$ observables can be measured with $\tilde{\mathcal{O}}(\sqrt{\mathcal{M}}/\varepsilon)$ preparations of the ground-state.
This approach proves beneficial when investigating defects or surface adsorption in periodic systems, where the focus is primarily on obtaining observables localised to the defect or adsorption site, rather than the bulk properties of the system.
If we utilise a more localised basis of size $\mathcal{M}$ to describe the defect or adsorption region, we have $\mathcal{M} \ll N \ll M$.
The basis could either be the real-space basis derived from the Fourier transform of the plane-wave basis, or the even smaller MO basis.
Consequently, we would prepare the ground state using the plane-wave basis in the first-quantization, apply basis transformation to the smaller basis localised around the defect or adsorption region (via \autoref{coro:basis_transformation}), and switch to the second quantization for the amplitude estimation algorithm~\cite{hugginsNearlyOptimalQuantum2022}.
The advantage of this technique is demonstrated in the second row of \autoref{tab:intro_table}.

\begin{figure}[t]
  \centering
  \includegraphics[width=\linewidth]{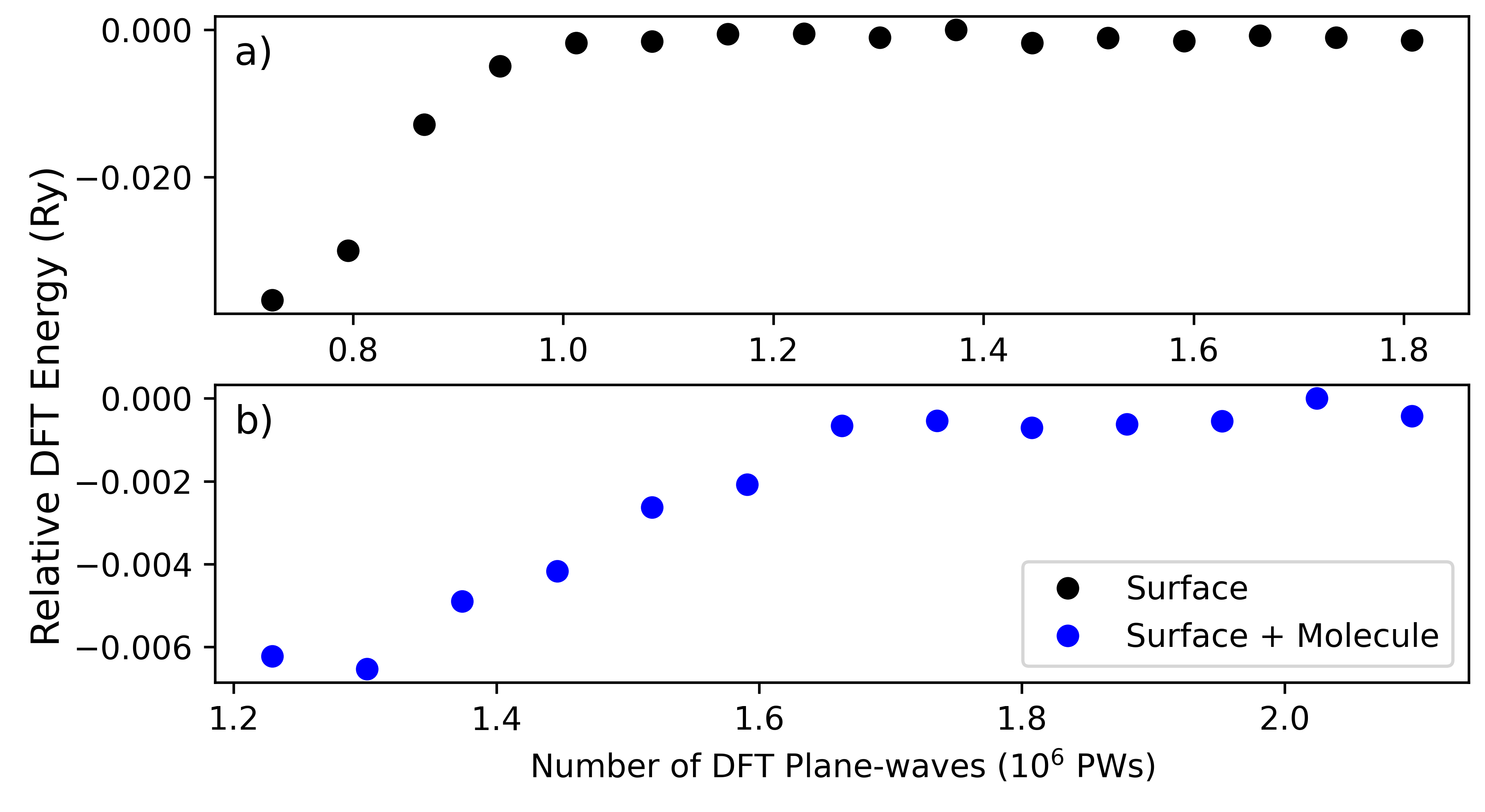}
  \caption{%
    The DFT energies of the \ce{Fe2O3} $(01\overline{1}2)$ surface.
    We calculated (a) the pristine surface and (b) the same surface with a benzene molecule, both for different simulation cell heights.
    The simulation details can be found in \siref{sec:fe2o3_surface_methods}{IV}.
  }
  \label{fig:fe2o3_benzene_plot}
\end{figure}

Aside from increasing the efficiency of observable measurements, the hybrid quantization framework could also be utilized to improve the ground-state preparation step in certain situations.
Consider the time-evolution calculation of a molecule as it is adsorbed onto a surface.
We initialise the system with the unentangled ground state of the molecule and the ground state of the surface.
Conventionally, we would place the molecule and the surface in the same simulation cell with a large vacuum space between the two to ensure no entanglement.
The ground state is then obtained in the first quantization using the plane-wave basis set.
However, the long-range nature of the Coulomb term would require a very large vacuum space, significantly increasing the number of plane waves.
As a rough example, DFT calculations of the \ce{Fe2O3} $(01\overline{1}2)$ surface with an unentangled benzene molecule, as shown in \autoref{fig:illustrations}(b), require approximately $\sim$\num{6.5E5} more plane waves compared to the surface alone (\autoref{fig:fe2o3_benzene_plot}).
As detailed in \siref{sec:fe2o3_surface_methods}{IV}, actual computational savings likely exceed this estimate as pseudopotentials are used for the DFT calculations.

\begin{table*}[!t]
  \centering
  \caption{%
    Summary of the complexity differences of Born-Oppenheimer AIMD for the MO basis with and without \autoref{thm:conversion}.
    For the row ``Electronic ground-state preparation'', $\alpha = |\langle\Psi_{\text{init}}|\Psi_g\rangle|$ represents the overlap between the initial state and the ground state~\cite{linNearoptimalGroundState2020,geFasterGroundState2019}.
    For the row ``Electronic forces calculation'', the quantum costs represents the number of state-preparations required.
    $N_{\text{ion}}$ represents the number of ions in the simulation.
  }
  \newcommand{\mrowlength}{3.5cm}
  \begin{tabularx}{\textwidth}{>{\raggedright}p{\mrowlength}p{5cm}L}
    \toprule
    Workflow steps&Without \autoref{thm:conversion}&With \autoref{thm:conversion}
    \\
    \midrule
    \multirow[t]{2}{\mrowlength}{Electronic ground-state preparation}
    &\multicolumn{2}{l}{\bfseries Quantum cost:}\\
    &$\tilde{\mathcal{O}}(M_{MO}^{2.1}/\alpha)$~\cite{leeEvenMoreEfficient2021} (2\textsuperscript{nd} quant)
    &$\tilde{\mathcal{O}}(M_{MO}^{2.1}/\alpha + N\log N\log M_{MO})$~\cite{leeEvenMoreEfficient2021} (2\textsuperscript{nd} quant $\rightarrow$ 1\textsuperscript{st} quant)
    \\
    \cmidrule(lr){1-3}
    \multirow[t]{4}{\mrowlength}{Electronic forces calculation}
    &\multicolumn{2}{l}{\bfseries Quantum cost (number of state preparations):}\\
    &$\tilde{\mathcal{O}}(M_{MO}^2/\varepsilon_{RDM})$~\cite{hugginsNearlyOptimalQuantum2022}
    &$\tilde{\mathcal{O}}(N^2 \log M_{MO}/\varepsilon_{RDM}^2)$~\cite{babbushQuantumSimulationExact2023}
    \\
    \addlinespace[2mm]
    &\multicolumn{2}{l}{\bfseries Classical cost:}\\
    &$\mathcal{O}(N_{\text{ion}}M_{MO}^4)$
    &$\mathcal{O}(N_{\text{ion}}M_{MO}^4)$
    \\
    \cmidrule(lr){1-3}
    \multirow[t]{2}{\mrowlength}{Ion simulation}
    &\multicolumn{2}{l}{\textbf{Classical costs:}}\\
    &$\mathcal{O}(N^2_{\text{ion}})$
    &$\mathcal{O}(N^2_{\text{ion}})$
    \\
    \cmidrule(lr){1-3}
    \multirow[t]{4}{\mrowlength}{Update electronic Hamiltonian}
    &\multicolumn{2}{l}{\bfseries Classical cost:}
    \\
    &$\mathcal{O}(M_{MO}^6)$
    &$\mathcal{O}(M_{MO}^6)$
    \\
    \bottomrule
  \end{tabularx}
  \label{tab:aimd_cost}
\end{table*}

This inefficiency can be avoided by computing the ground states of the molecule and surface as isolated systems, followed by their tensor product.
The molecule's ground state may even be calculated in the MO basis, which requires transformation to the plane-wave basis (per \autoref{coro:basis_transformation}) to align with the surface's plane-wave basis before applying the tensor product.
The resulting wavefunction is then evolved under the Hamiltonian of the combined system.

Applying the tensor product naively in the sorted-list encoding violates its ascending-order requirement and may introduce states with duplicate orbital indices, necessitating their removal from the wavefunction.
Similarly, in the first quantization, the tensor product breaks the antisymmetric property.
As a result, this tensor product represents a novel application of the hybrid quantization scheme, as both the first and second quantization face significant limitations in their implementation.
Unlike the other applications, where the suboptimal quantization may suffice, this application requires the conversion circuit.
To address these challenges, we would first perform the tensor product in the second-quantized sorted-list encoding by concatenating the qubit representation of the two systems.
A sorting network then restores the ascending-order requirement, with $Z$ gates applied for each swap to account for the parity.
The wavefunction is subsequently converted to the first quantization, with the states containing duplicate orbital indices flagged by an ancilla qubit using $\mathcal{O}(N)$ comparisons, yielding the tensor product in the first quantization.

Another application involves identifying active sites in enzymes or catalysts, where we need to perform multiple ground-state calculations of the reactant-catalyst systems across various atomic configurations.
Without the hybrid quantization framework, those ground-state calculations of multiple different atomic configurations are independent calculations.
Thus, we would incur a cost of $\tilde{\mathcal{O}}(\lambda/\alpha)$~\cite{linNearoptimalGroundState2020,geFasterGroundState2019} applications of the qubitization circuits for every atomic configurations, with $\lambda$ representing the 1-norm of the Hamiltonian and $\alpha = |\langle\Psi_{\text{init}}|\Psi_g\rangle|$ representing the absolute value of the overlap between the initial state and the ground state.
With the hybrid quantization framework, the ground states of the catalyst and the reactants are first prepared independently and their tensor product are used as effective initial states for the ground-state simulation of the combined system, reducing the costs involved with preparing good initial states for the ground-state preparation algorithm.
For enzymes or zero-dimensional catalysts, both components are represented in the MO basis.
For two-dimensional catalysts, the reactant remains in the MO basis, while the catalyst is described using the plane-wave basis, as illustrated in \autoref{fig:material_characterization}(b).
To proceed, both ground states are transformed to the basis of the combined system using \autoref{coro:basis_transformation}, followed by the application of the tensor product.
Here, we would use the procedure outlined in the previous example to perform the tensor product.

\medskip
\noindent\textbf{Born-Oppenheimer AIMD simulation.}
This example focuses on the application of \autoref{thm:conversion} for Born-Oppenheimer AIMD (BO-AIMD)~\cite{bornZurQuantentheorieMolekeln1927} simulations using the MO basis.
While previous works has suggested that the plane-wave basis offers better asymptotic complexity with respect to $M$~\cite{babbushQuantumSimulationExact2023,obrienEfficientQuantumComputation2022}, calculations in the MO basis remains valuable in several context.
This includes multi-scale simulations like QM/MM~\cite{warshelTheoreticalStudiesEnzymic1976,sennQMMMMethods2009} and quantum embedding~\cite{weisburnMultiscaleEmbeddingQuantum2025,caoInitioQuantumSimulation2023,liuBootstrapEmbeddingQuantum2023,rossmannekQuantumEmbeddingMethod2023,shangPracticalMassivelyParallel2023}.
In these two cases, the effective Hamiltonians are almost always provided in an arbitrary basis.
Furthermore, embedding techniques which splits the system into fragments also may not have well defined simulation cell, hindering the use of the plane-wave basis.

\begin{figure}[t]
  \centering
  \includegraphics[width=\linewidth]{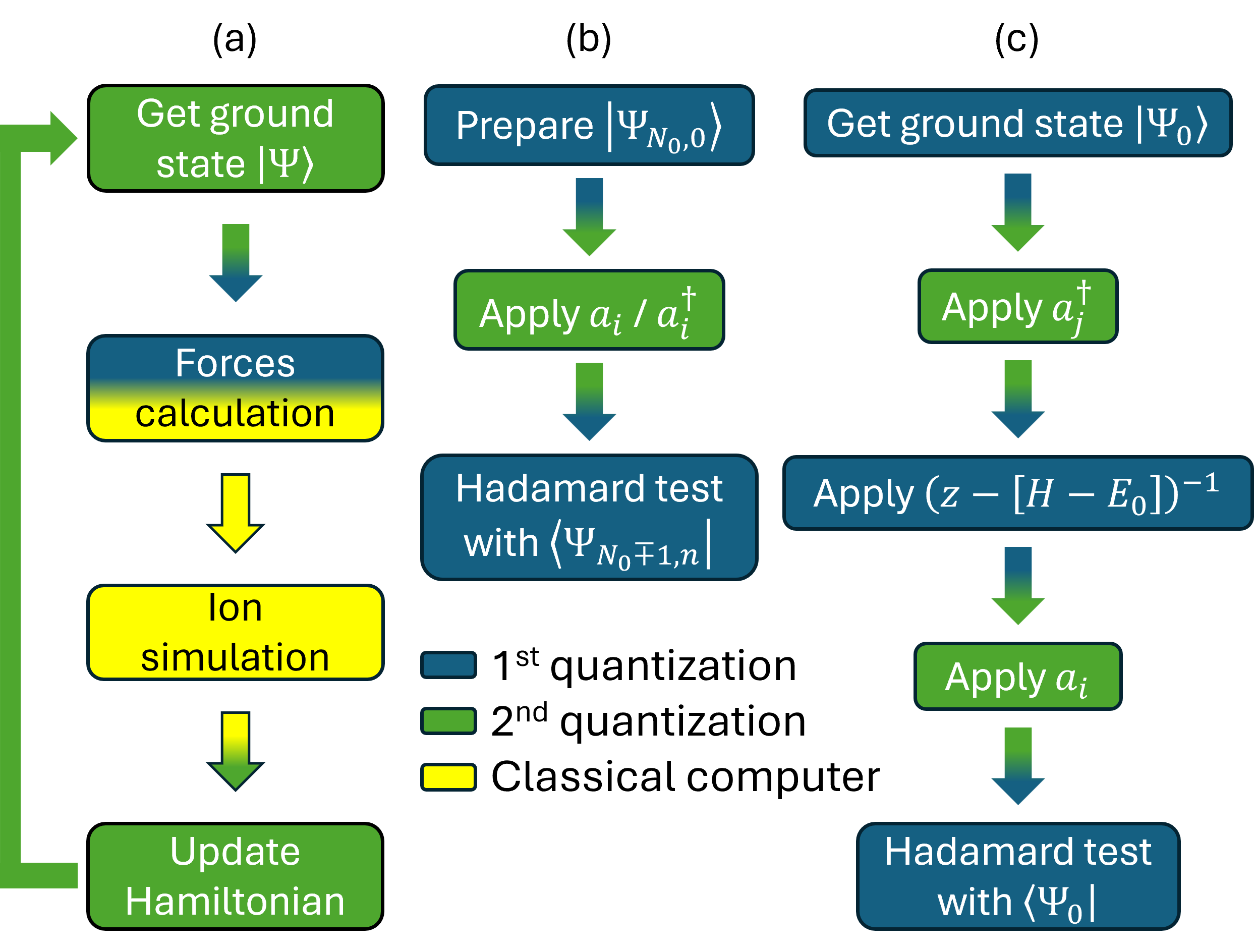}
  \caption{
    Flowcharts to illustrate the workflow of AIMD and characterization of the spectroscopic properties of materials.
    (a) BO-AIMD simulations MO basis to extract RDMs for the calculation of atomic forces.
    (b) Measurement of ionization and electron attachment probabilities of \autoref{eqn:ion_att_probs}.
    (c) Measurement of the Green's function of \autoref{eqn:greens_2}.
  }
  \label{fig:basis_transform}
\end{figure}

Due to the Born-Oppenheimer approximation, the electronic degrees of freedom are simulated in quantum computers, while the nuclear degrees of freedom are simulated in classical computers.
As a result, a typical BO-AIMD simulation workflow comprises four iterative steps: electronic ground-state preparation, electronic forces calculations, ion simulation, and finally updating the electronic Hamiltonian, illustrated in \autoref{fig:basis_transform}(a) and with the costs summarized by the four rows of \autoref{tab:aimd_cost}.
These steps are repeated for a predetermined number of time steps.

First, the ground-state is prepared using $\tilde{\mathcal{O}}(\lambda/\alpha)$ applications of the qubitization circuits~\cite{linNearoptimalGroundState2020,geFasterGroundState2019}.
Next, forces exerted by the electrons acting on the ions are calculated.
Using Hellmann-Feynman theorem~\cite{feynmanForcesMolecules08,laiAccurateEfficientCalculations2023}, the forces can be written as the ground-state expectation values of the Hamiltonian's first derivative with respect to the ionic positions $\vb{R}_I$ as:
\begin{align*}
  \vb{F} = -\ev**{\dv{H}{\vb{R}_I}}{\Psi} =& -\sum_{pq}\dv{h_{pq}}{\vb{R}_I}\ev**{a_p^\dagger a_q}{\Psi}\\
         &- \sum_{pqrs}\dv{h_{pqrs}}{\vb{R}_I}\ev**{a_p^\dagger a_q^\dagger a_r a_s}{\Psi}.
\end{align*}
This requires measurement of the 1- and 2-RDMs along with a classical computations with costs of $\mathcal{O}(N_{\text{ion}}M_{MO}^4)$, where $N_{\text{ion}}$ represents the number of ions.
In the case where the first-quantized classical shadows measurement algorithm is more efficient, \autoref{thm:conversion} could be used to convert the ground-state to the first-quantization.
The third step is the simulation of the ions, where the forces obtained from the previous step is used to move the ions using classical mechanics, which should scale by $\mathcal{O}(N_{\text{ion}}^2)$.

Finally, the electronic Hamiltonian is updated to reflect the new positions of the ions.
For BO-AIMD calculations with the MO basis, additional complexity arises in this step from the dependence of basis functions on atomic positions as shown in the fourth row of \autoref{tab:aimd_cost}.
In this case, the $h_{pq}$ and $h_{pqrs}$ integrals must be recalculated at every time step, with classical costs of $\mathcal{O}(M^6)$ incurred when constructing efficient qubitization circuits using rank-reduction techniques, such as single-factorization~\cite{berryQubitizationArbitraryBasis2019}, double-factorization~\cite{vonburgQuantumComputingEnhanced2021}, and tensor hypercontraction~\cite{leeEvenMoreEfficient2021}.
This incurs a classical cost of $\mathcal{O}(M_{MO}^2)$ for calculating the basis transformation matrix.

\begin{figure}[t]
  \centering
  \includegraphics[width=\linewidth]{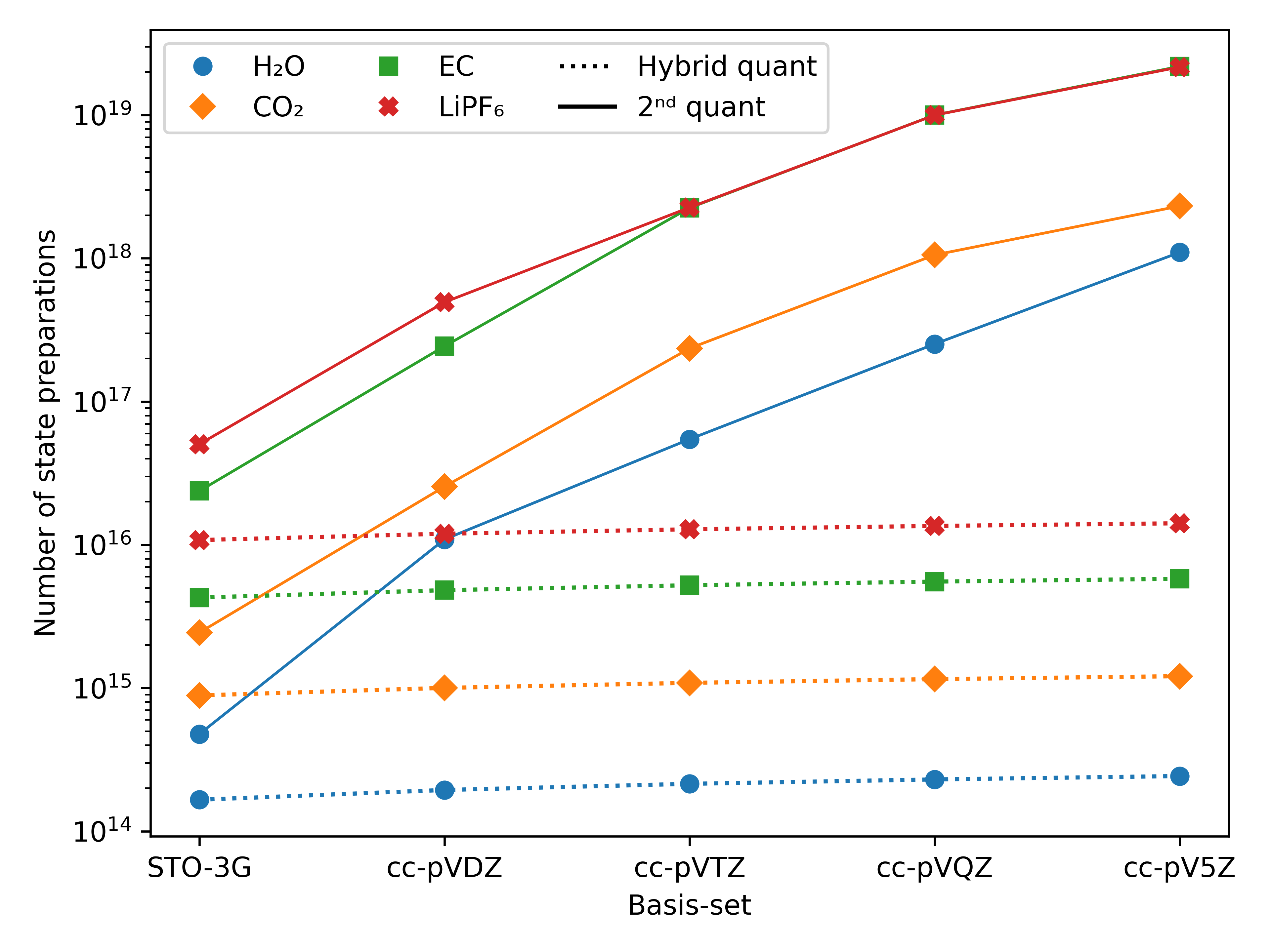}
  \caption{%
    Quantitative estimates for the number of state preparations required to obtain the 2-RDM.
    Each of the 2-RDM elements is calculated to a precision of $10^{-3}$ Ha, with a success probability of $99\%$.
    The molecular systems used in the test includes the \ce{H2O}, \ce{CO2}, ethylene carbonate \ce{(CH2O)2CO} (EC), and \ce{LiPF6}, and standard basis-sets including the STO-3G, cc-pVDZ, cc-pVTZ, cc-pVQZ, and cc-pV5Z bases are included
  }
  \label{fig:aimd_plot}
\end{figure}

To provide a quantitative estimation on the cost reduction of \autoref{thm:conversion}, we obtained estimates for the number of state preparations required for typical small molecular systems including \ce{H2O}, \ce{CO2}, ethylene carbonate \ce{(CH2O)2CO} (EC), and \ce{LiPF6}.
\autoref{fig:aimd_plot} shows the comparison between the first-quantized classical shadows method (labelled as ``hybrid quant'') and the second-quantized gradient-based method (labelled as ``2\textsuperscript{nd} quant'').
For this case, the first-quantized classical shadows method, which requires the use of the hybrid quantization scheme, consistently demonstrates an advantage over the second-quantized method across all the included basis-sets.
This advantage ranges from $\sim$3 times fewer for the STO-3G basis-set to around 3 orders of magnitude fewer applications of the ground-state for the cc-pV5Z basis-set.
Thus, the implementation of our hybrid quantization scheme offers a compelling pathway to accelerate BO-AIMD simulations, yielding orders of magnitude improvement in computational cost for force evaluations.

\medskip
\noindent\textbf{Spectroscopic, electron ionization, and attachment characterization.}
For this example, we focus on the study of spectroscopic properties, namely the calculation of the Green's function, along with the study of electron ionization and attachment processes of periodic systems with the plane-wave basis-set.
The plane-wave basis-set favours the first-quantized Hamiltonian simulation algorithm for efficient computation.
However, those processes are electron non-conserving which necessitates the second-quantized representation.
An example of this is the ionization and electron attachment probabilities~\cite{runggerDynamicalMeanField2020}, written as
\begin{equation}
  \begin{aligned}
    \lambda_{h,i,n} &= |\mel{\Psi_{N_0-1,n}}{a_i}{\Psi_{N_0,0}}|^2\text{  and}\\
    \lambda_{p,i,n} &= |\mel{\Psi_{N_0+1,n}}{a_i^\dagger}{\Psi_{N_0,0}}|^2,
  \end{aligned}
  \label{eqn:ion_att_probs}
\end{equation}
respectively.
In this case, $N_0$ denotes the electron count for a neutral systems, $\ket{\Psi_{N,n}}$ is the $n$-th excited state of the $N$-electron system, with $n=0$ being the ground state.
To calculate the above probabilities, we are forced to use the suboptimal second-quantized plane-wave Hamiltonian simulation algorithm as $a_i$ and $a_i^\dagger$ are electron non-conserving.
With the hybrid quantization framework, we could prepare $\ket{\Psi_{N_0,0}}$ and $\ket{\Psi_{N_0\mp 1,n}}$ in the first quantization and apply $a_i$ or $a_i^\dagger$ in the second quantization.
The Hadamard test workflow, depicted in \autoref{fig:basis_transform}(b), would then involve the preparation of $\ket{\Psi_{N_0,0}}$, conversion to second quantization, application of $a_i$ or $a_i^\dagger$, conversion back to first quantization, and finally the conjugate preparation of $\ket{\Psi_{N_0\mp 1,n}}$.
Since the conversion circuit costs $\mathcal{O}(N\log N\log M)$ gates and the application of $a_i$ or $a_i^\dagger$ in the sorted-list encoding costs $\mathcal{O}(N\log M)$ gates, the complexity of this whole procedure is dominated by the preparation of $\ket{\Psi_{N_0,0}}$ and $\ket{\Psi_{N_0\mp 1,n}}$.
The ground-state preparations costs $\tilde{\mathcal{O}}(\lambda/\alpha)$ queries to the qubitization circuits.
Thus, the Toffoli scaling advantage of the hybrid quantization here is equal to the Toffoli scaling advantage of the first-quantized plane-wave Hamiltonian simulation.

Another example is the calculation of the Green's function, essential in calculating the optical properties of materials, performing computational spectroscopy, and studying electron transport and scattering~\cite{negeleQuantumManyparticleSystems1988,martinInteractingElectronsTheory2016}.
In fact, the ionization and electron attachment probabilities above contributes to the Green's function in the Lehmann representation~\cite{runggerDynamicalMeanField2020}, where $\delta$ below is an infinitesimally small positive number
\begin{align}
  G_{\alpha}(\omega) &= \sum_{n=0}^{M}\frac{\lambda_{h,\alpha,n}}{\omega+i\delta-(E_0 - E_{N_0-1,n})} \notag\\
                     &\phantom{=} + \sum_{n=0}^{M}\frac{\lambda_{p,\alpha,n}}{\omega + i\delta - (E_{N_0+1,n}-E_0)}.\label{eqn:greens_1}
\end{align}

Another representation of the Green's function discussed in the literature~\cite{tongFastInversionPreconditioned2021}, which only require the preparation of the ground state, is the following:
\begin{equation}
  \begin{aligned}
    G^{(+)}_{ij}(z) &= \ev{a_i(z-[H - E_0])^{-1} a_j^\dagger}{\Psi_{N_0,0}},\\
    G^{(-)}_{ij}(z) &= \ev{a_j^\dagger(z+[H - E_0])^{-1} a_i}{\Psi_{N_0,0}}.
  \end{aligned}
  \label{eqn:greens_2}
\end{equation}
Similarly with the ionization and electron attachment probabilities, we would use the first quantization for optimal calculations of both the ground state $\ket{\Psi_0}$ and the application of the $(z\pm[H-E_0])^{-1}$ term, while performing the electron non-conserving $a_i$ and $a_j^\dagger$ parts in the second-quantized representation, with the workflow shown in \autoref{fig:basis_transform}(c).
In this case, the complexity of this procedure is dominated by the application of the $(z\pm[H-E_0])^{-1}$ term.

Using the preconditioned fast inversion algorithm~\cite{tongFastInversionPreconditioned2021}, we can obtain a block-encoding of $(z\pm[H-E_0])^{-1}$ for some $\eta \ge |z|$ using $\tilde{\mathcal{O}}(N^6M\ln(\delta^{-1})/\Omega\eta^2\varepsilon)$ applications of the first-quantized qubitization algorithm (details in \siref{sec:greens_methods}{V}), where $\Omega$ represents the simulation cell volume, offering a significant improvement in the plane-wave basis where $N\ll M$.
For comparison, the second-quantized Jordan-Wigner encoding using the plane-wave dual basis scales as
$\tilde{\mathcal{O}}(M^5\ln(\delta^{-1})/\Omega^2\eta^2\varepsilon)$~\cite{tongFastInversionPreconditioned2021}.
Alternatively, we could also use the non-preconditioned inversion method~\cite{childsQuantumAlgorithmSystems2017}.
In this case, the block-encoding of $(z\pm[H-E_0])^{-1}$ is achieved using $\tilde{\mathcal{O}}(NM^{2/3}/\Omega^{2/3}\eta^2\varepsilon + N^2M^{1/3}/\Omega^{1/3}\eta^2\varepsilon)$ applications of the first-quantized Hamiltonian (details in \siref{sec:greens_methods}{V}), compared to $\tilde{\mathcal{O}}(M^{7/3}/\Omega^{2/3}\eta^2\varepsilon)$ applications for the second-quantized Jordan-Wigner encoding~\cite{tongFastInversionPreconditioned2021}, as summarized in the third row of \autoref{tab:intro_table}.

\medskip
\noindent\textbf{\large Discussion}\\
In this study, we present a hybrid quantization scheme that significantly enhances material simulations by enabling efficient transitions between first- and second-quantized encodings.
This approach achieves polynomial improvements through an efficient $\mathcal{O}(N\log N\log M)$ conversion circuit by using the most efficient quantization for each simulation step.
The conversion circuit, central to this scheme, incurs a gate cost of $\mathcal{O}(N\log N\log M)$ for a system of $N$ electrons and $M$ orbitals, a cost that is negligible compared to the significant savings provided by its application.
This scheme also mitigates limitations inherent to both encodings, such as applying particle non-conserved operators in the first quantization, while preserving the optimal efficiency of plane-wave and MO Hamiltonian simulations.
Additionally, the hybrid quantization scheme facilitates the combination of quantum states prepared in distinct bases, addressing challenges in tensor product operations of isolated wavefunctions.
We also introduced a basis transformation algorithm for the first-quantized encoding, allowing quantum simulations to efficiently switch between the plane-wave and MO bases.

Prior work has utilized the antisymmetrization circuit~\cite{berryImprovedTechniquesPreparing2018} to convert the Jordan-Wigner encoding to the first-quantized encoding for efficient $k$-RDM measurements~\cite{babbushQuantumSimulationExact2023}.
Our hybrid quantization scheme improves on this by introducing the reverse conversion.
Furthermore, the sorted-list encoding~\cite{carolanSuccinctFermionData2024} is used instead for the second-quantized counterpart, greatly simplifying the conversion and reducing the qubit requirement of both quantizations to $\mathcal{O}(N\log M)$.
This, in turn, also opens up the sorted-list encoding to additional applications, especially in encoding electron non-conserving operations that is applied to first-quantized wavefunctions.

As our hybrid quantization framework uses the sorted-list encoding for the second-quantized representation, future research should focus on implementing and benchmarking this encoding in various Hamiltonian simulation algorithms, such as trotterization and qubitization.
While our conversion circuit converts between the first-quantized encoding and the sorted-list encoding, the principles extend to other potential conversion circuits.
Hence, exploring alternative circuits for different first- and second-quantized encodings remain a compelling avenue for further investigation.

\medskip
\noindent\textbf{\large Methods}
\begin{proof}[Proof of Theorem 1]
  The extended proof can be found in the \hyperlink{sec:thm_1_extended_proof}{Supplementary Methods}.
  Here, we show a high-level proof of the conversion circuit.
  The conversion from the sorted-list encoding to the first-quantized encoding can be performed using the antisymmetrization circuit~\cite{berryImprovedTechniquesPreparing2018}.
  In total, they reported a gate cost of $\mathcal{O}(N\log N\log M)$ and an ancilla cost of $\mathcal{O}(N\log N)$.
  The gate cost here is dominated by the implementation of a reverse sorting network which applies a $Z$ gate when any two registers are swapped.

  On the other hand, the conversion from the first-quantized encoding to the sorted-list encoding is simpler.
  Given an antisymmetrized first-quantized state, we would simply sort the states in an ascending order through the use of a sorting network shown in the following equation
  \begin{gather*}
    \sum_{\sigma\in S_N}\operatorname{sgn}(\sigma)\ket{i_{\sigma(1)}}\otimes\ket{i_{\sigma(2)}}\otimes\cdots\otimes\ket{i_{\sigma(N)}}\otimes\ket{0}\nonumber\\
    \xRightarrow{\text{sort}}\ket{i_1}\otimes\ket{i_2}\otimes\cdots\otimes\ket{i_N},
  \end{gather*}
  with $i_1 < i_2 < \ldots < i_N$.
  Here, the left hand side represents the first-quantized representation of a Slater determinant with $S_N$ representing the symmetric group of $N$ symbols, and $\operatorname{sgn}(\sigma)$ denotes the sign of the permutation.
  Similarly with the antisymmetrization circuit, a $Z$ gate is also applied when any two registers are swapped to account for the $\operatorname{sgn}(\sigma)$ term.
  The sorted-list state shown on the right-hand side above is obtained after the registers are sorted.
  As a result, the conversion from the first-quantized encoding to the sorted-list encoding also costs $\mathcal{O}(N\log N\log M)$ gates and $\mathcal{O}(N\log N)$ ancillae.
\end{proof}

\begin{proof}[Proof of Corollary 1]
  Basis transformations of the single-particle wavefunctions can be represented with an $M_1\times M_2$ matrix
  \begin{equation}
    \psi^{(1)}_i(\vb{x}) = \sum_j U_{ij}\psi^{(2)}_j(\vb{x}).\label{eqn:basis_transformation_fq}
  \end{equation}
  Thus, the basis transformation of a Hartree product can be performed by applying the matrix $U$ to each electronic degrees of freedom
  \begin{equation*}
    \Psi = \left( \sum_{j_1}U_{i_1j_1}\psi_{j_1}^{(2)}(\vb{x}_1) \right)\otimes\cdots\nonumber \otimes\left( \sum_{j_N}U_{i_Nj_N}\psi_{j_N}^{(2)}(\vb{x}_N) \right).
  \end{equation*}

  In a quantum circuit, this basis transformation involves the circuit implementing $U$, applied to each of the $N$ registers independently
  \begin{equation*}
    \ket{i_1}\otimes\cdots\otimes\ket{i_N}\xRightarrow{U} (U\ket{i_1})\otimes\cdots\otimes(U\ket{i_N}).
  \end{equation*}
  Assuming the worst-case complexity of an arbitrary dense matrix, $U$ could be implemented using single-qubit gates and $\mathcal{O}(M_1M_2)$ CNOT gates~\cite{itenQuantumCircuitsIsometries2016,malvettiQuantumCircuitsSparse2021}.
  With $N$ electrons, the total complexity is $\mathcal{O}(NM_1M_2)$.
\end{proof}

\medskip
\noindent\textbf{ACKNOWLEDGEMENTS}\\
A.H. gratefully acknowledges the sponsorship from City University of Hong Kong (Project No. 7005615, 7006103), CityU Seed Fund in Microelectronics (Project No. 9229135), National Natural Science Foundation of China (NSFC) (Grant No. 62541160274), and Hon Hai Research Institute (Project No. 9231594).
This work was carried out using the computational facilities, CityU Burgundy, managed and provided by the Computing Services Centre at City University of Hong Kong (\url{https://www.cityu.edu.hk}).

\medskip
\noindent\textbf{AUTHOR CONTRIBUTIONS}\\
The project was conceived by C.K. and Y.C.
Theoretical results were proved by C.K. in discussion with M.H.
Numerical simulations and analysis were performed by C.K. in discussion with A.H.
All authors contributed to the write-up.

\medskip
\noindent\textbf{COMPETING INTERESTS}\\
The authors declare no competing interests.

\medskip
\noindent\textbf{References}\\
\bibliography{main}

\newpage

\clearpage

\onecolumngrid

\appendix


\renewcommand{\figureautorefname}{Supplementary Figure}
\renewcommand{\figurename}{Supplementary Figure}
\renewcommand{\thefigure}{\arabic{figure}}

\renewcommand{\tableautorefname}{Supplementary Table}
\renewcommand{\tablename}{Supplementary Table}
\renewcommand{\thetable}{\arabic{table}}
\renewcommand{\equationautorefname}{Supplementary Equation}
\renewcommand\theequation{\arabic{equation}}

\renewcommand{\appendixname}{}
\renewcommand\thesection{Supplementary Note \arabic{section}}
\renewcommand\thesubsection{Supplementary Note \arabic{section}.\arabic{subsection}}

\begin{center}
\Large Supporting Information for ``Optimizing Quantum Chemistry Simulations with a Hybrid Quantization Scheme''
\end{center}

\section{Summary of mentioned encodings}
\label{sec:prev_encoding}
\subsection{First-Quantized Encoding}
\label{sec:prev_encoding_fq}
The first-quantized representation uses the tensor products of the single-particle wavefunctions $\psi_1,\ldots,\psi_N$, called Hartree products, as a basis.
Since a single Hartree product does not obey the Pauli exclusion principle, fermionic antisymmetry is enforced by explicitly expressing the wavefunction as a linear combination of Hartree products, forming a Slater determinant.
This can be represented as a matrix determinant with $N!$ terms.
\begin{align*}
  [\psi_1,\ldots,\psi_N] &= \frac{1}{\sqrt{N!}}\left|
  \begin{matrix}
    \psi_1(\vb{x}_1)&\cdots&\psi_N(\vb{x}_1)\\
    \vdots&\ddots&\vdots\\
    \psi_1(\vb{x}_N)&\cdots&\psi_N(\vb{x}_N)
  \end{matrix}
  \right|
\end{align*}
Slater determinants serve as the basis for the second-quantized representation, while the Hartree products are used as the basis for the first-quantized representation.
The first-quantized encoding~\cite{suFaultTolerantQuantumSimulations2021,babbushQuantumSimulationChemistry2019} encodes a single Hartree product as follows.
\begin{align*}
  \Psi &= \psi_{i_1}(\vb{x}_1)\otimes\psi_{i_2}(\vb{x}_2)\otimes\cdots\otimes\psi_{i_n}(\vb{x}_1)\\
  \mathcal{E}(\Psi) &= \ket{i_1}\otimes\ket{i_2}\otimes\cdots\otimes\ket{i_n}
\end{align*}
Here, $\ket{i_1},\ldots,\ket{i_n} \in \{\ket{1}, \ldots, \ket{M}\}$ are binary representations of indices to the single-particle basis $\ket{\psi_1(\vb{x})},\ldots,\ket{\psi_M(\vb{x})}$,
each encoded in $\lceil \log_2 M\rceil$ qubits for $N$ electrons, totalling $N\lceil\log_2 M\rceil$ qubits. Unlike the sorted-list encoding, the indices are not sorted, as different orderings correspond to distinct Hartree products.
The antisymmetrization of the wavefunction here must be performed explicitly during the quantum simulation process, primarily during state initialization~\cite{babbushQuantumSimulationChemistry2019}.
Hamiltonian simulation preserves this antisymmetry as long as the input wavefunction is symmetric, eliminating the need for continuous antisymmetrization.

\subsection{Sorted-List Encoding}
\label{sec:prev_encoding_sl}
The sorted-list encoding is first proposed by \citett{carolanSuccinctFermionData2024}.
While they introduced several variants of the $\mathcal{O}(N\log M)$ second-quantized encodings, we only used the variant detailed in Section 4.2 of their paper, which we termed the sorted-list encoding.
This variant has the simplest encoding along with the most efficient circuit implementations in terms of gate count, while others increase the complexity of the circuit implementations to save either the qubit count and/or the circuit depth.
This variant also allows us to create efficient conversions between the first-quantized and second-quantized encodings.

The sorted-list encoding also has comparable asymptotic gate complexity scaling with the Jordan-Wigner encoding for most Hamiltonian simulations.
More specifically, for trotterization-based circuits, the time evolution of fermionic operators ($\exp(i\theta (O + O^\dagger))$, with $O=a_p^\dagger a_q, ia_p^\dagger a_q, a_p^\dagger a_q^\dagger a_r a_s, ia_p^\dagger a_q^\dagger a_ra_s, \ldots$) incurs a gate cost of $\mathcal{O}(N\log M)$, as shown in Section 2.3 of \citett{carolanSuccinctFermionData2024}, compared to $\mathcal{O}(M)$ for the Jordan-Wigner encoding.
Similarly, the SELECT subroutine used in qubitization-based algorithms can also be implemented with $\mathcal{O}(N\log M)$ gates, as shown in Theorem 8.3 of \citett{carolanSuccinctFermionData2024}, compared to $\mathcal{O}(M)$ gates of the Jordan-Wigner encoding.

Let $\vb{x}\in\mathbb{F}_2^M$ be the bitstring representing a Slater determinant with Hamming weight $|\vb{x}| = N$, indicating $N$ electrons.
We then define $\vb{e}_i\in\mathbb{F}_2^M$ to be a bitstring with a `1' in position $i$ and `0' elsewhere. Thus, $\vb{x} = \vb{e}_{i_1}\oplus\vb{e}_{i_2}\oplus\cdots\oplus\vb{e}_{i_N}$ with $i_1 < i_2 < \cdots < i_N$.
The sorted-list encoding $\mathcal{E}$ concatenates the binary representation of each of the occupied orbital indices in ascending order.
\begin{equation*}
  \mathcal{E}(\vb{x}) = \ket{i_1}\otimes\ket{i_2}\otimes\cdots\otimes\ket{i_N}\otimes\ket{\infty}\otimes\cdots\otimes\ket{\infty}
\end{equation*}
Each orbital index $\ket{i}$ is encoded as a binary string of $\lceil\log_2 M\rceil$ qubits.
The number of registers $N_{\text{reg}}$ does not need to equal to the number of electrons $N$.
To support particle non-conserving operations, a sentinel state, representing an ``unoccupied'' orbital, is used to fill in the empty registers when $N < N_{\text{reg}}$. By convention, the sentinel state is represented by the symbol $\ket{\infty}$ and is ordered greater than all orbital indices.
Thus, $N_{\text{reg}}\lceil\log_2 (M+1)\rceil$ qubits can hold all bitstrings with Hamming weight less than or equal to $N_{\text{reg}}$.
This enables the sorted-list encoding to perform particle non-conserving operations as long as enough registers are set aside for the operation.

\begin{figure}[!h]
  \centering
  \includegraphics{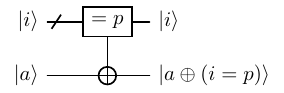}
  \caption{The $=p$ circuit~\cite{carolanSuccinctFermionData2024}. This circuit flips $t$ when the value of $i$ is equal to a constant $p$.}
  \label{fig:eq_const_circuit}
\end{figure}

\begin{figure}[!h]
  \centering
  \includegraphics{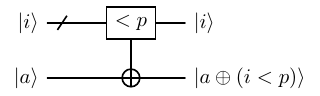}
  \caption{The $<p$ circuit~\cite{carolanSuccinctFermionData2024}. This circuit flips $t$ when the value of $a$ is less than a constant $p$. While the figure above shows the $<p$ gate, the original work includes the $\le p$ and $>p$ variant which can be implemented similarly~\cite{berryQubitizationArbitraryBasis2019}.}
  \label{fig:lt_const_circuit}
\end{figure}

\begin{figure}[!h]
  \centering
  \includegraphics{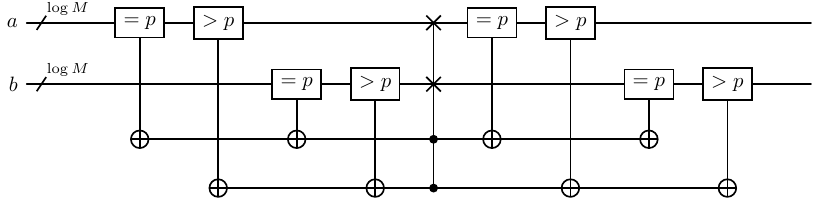}
  \caption{The bubble circuit $U_p$~\cite{carolanSuccinctFermionData2024}. This circuit swaps the two registers $a$ and $b$ when one of them is equal to $p$ and the other is larger than $p$. The $=p$ circuit is simply the $\log M$ qubit Toffoli gate while the $>p$ gate can be adapted from \autoref{fig:lt_const_circuit}.}
  \label{fig:bubble_ori_circuit}
\end{figure}

\begin{figure}[!h]
  \centering
  \includegraphics{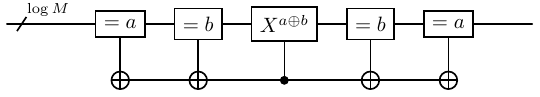}
  \caption{The $a\leftrightarrow b$ circuit~\cite{carolanSuccinctFermionData2024}. This circuit outputs $b$ when the input is $a$ and outputs $a$ when the input is $b$. Otherwise, the input is unchanged. The $X^{a\oplus b}$ is composed of multiple CNOT gate according to the binary representation of $a\oplus b$.}
  \label{fig:swap_circuit}
\end{figure}

Before elaborating how fermionic operations are encoded, we first introduce several building block gates frequently used in the sorted-list encoding.
The $=p$ gate (\autoref{fig:eq_const_circuit}) accepts one register as input and flips a target ancilla qubit if that register is equal to $p$.
Similarly, the $<p$ gate (\autoref{fig:lt_const_circuit}) accepts one register as input and flips a target ancilla qubit if that register is less than $p$.
The bubble gate $U_p$ (\autoref{fig:bubble_ori_circuit}) accepts two registers as input and swaps the two registers if one of them is equal to $p$ and the other is larger than $p$.
Finally, the $a\leftrightarrow b$ (\autoref{fig:swap_circuit}) gate outputs $b$ when the input is $a$ and outputs $a$ when the input is $b$. Otherwise the input remains unchanged.

\begin{figure}[!h]
  \centering
  \includegraphics{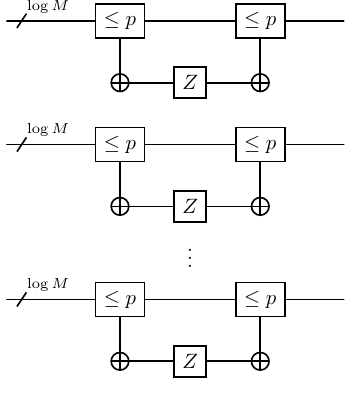}
  \caption{The $\operatorname{sgn-rank}(p)$ circuit~\cite{carolanSuccinctFermionData2024}. The $\le p$ circuit is shown on \autoref{fig:lt_const_circuit}. Quantum wires not extending to both ends of the circuit are ancilla qubits.}
  \label{fig:sgnrank_circuit}
\end{figure}

\begin{figure}[!h]
  \centering
  \includegraphics{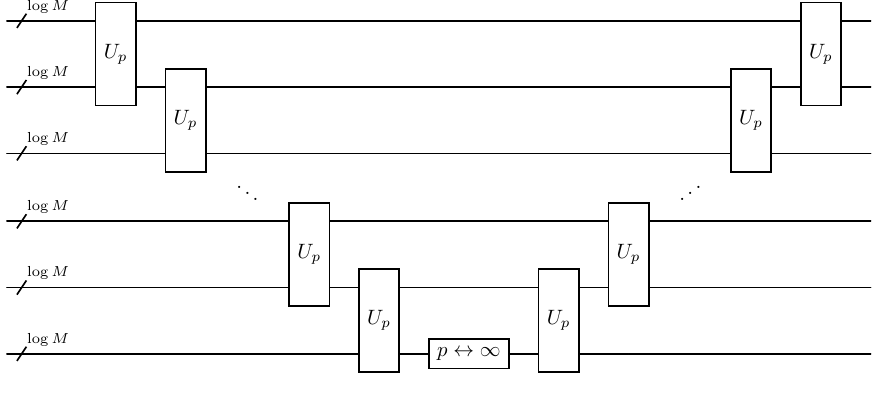}
  \caption{The $\operatorname{bit-flip}(p)$ circuit~\cite{carolanSuccinctFermionData2024}. The bubble circuit $U_p$ is shown on \autoref{fig:bubble_ori_circuit}, while the swap circuit $p\leftrightarrow\infty$ is shown in \autoref{fig:swap_circuit}.}
  \label{fig:bitflip_circuit}
\end{figure}

Fermionic operations on this encoding are first decomposed to Majorana operators.
The Majorana operators are then further decomposed to $\operatorname{sgn-rank}$ and $\operatorname{bit-flip}$ operators.
\begin{align*}
  a_p^\dagger &= \frac{1}{2}(\gamma_{2p-1}-i\gamma_{2p}) & a_p &= \frac{1}{2}(\gamma_{2p-1}+i\gamma_{2p})\\
  \mathcal{E}(\gamma_{2p-1}) &= \operatorname{bit-flip}(p)\operatorname{sgn-rank}(p-1) & \mathcal{E}(\gamma_{2p}) &= i\operatorname{bit-flip}(p)\operatorname{sgn-rank}(p)
\end{align*}
The $\operatorname{sgn-rank}(p)$ operator calculates the parity of the number of occupied states for orbital indices less than or equal to $p$. It then multiplies the wavefunction by $-1$ for odd parity and $+1$ for even parity. On the other hand, the $\operatorname{bit-flip}(p)$ operator flips the occupation of orbital $p$.
The circuit implementations of the $\operatorname{sgn-rank}$ and $\operatorname{bit-flip}$ operators can be found in \autoref{fig:sgnrank_circuit} and~\ref{fig:bitflip_circuit}, respectively.
Essentially, the $\operatorname{sgn-rank}(j)$ circuit applies a $Z$ gate for every register containing an orbital index less than or equal to $j$.
On the other hand, $\operatorname{bit-flip}(j)$ first checks if $j$ is occupied.
If $j$ is occupied it replaces $\ket{j}$ with $\ket{\infty}$, otherwise it replaces $\ket{\infty}$ with $\ket{j}$.
After replacing the target register with the appropriate value, that register is sorted back to the correct ascending order.
As the Majorana operators changes the number of occupied orbitals, the encoding needs to set aside more registers than the total number of electrons.
More specifically, the one-electron terms will require 2 extra registers in the encoding than the total number of electrons, while the two-electron terms will require 4 extra registers.
Such registers are filled with the $\ket{\infty}$ state so that the swap circuit in \autoref{fig:bitflip_circuit} can be done.

\section{Number of state preparations applications for \texorpdfstring{$k$}{k}-RDM measurements}
\label{sec:rdm_formula}
The number of applications of the state preparation for the first-quantized classical shadows algorithm~\cite{babbushQuantumSimulationExact2023} is shown in Equation~36 of the SI of the paper~\cite{babbushQuantumSimulationExact2023}.
In this case, the number of measurements to such that each $k$-RDM can be estimated to precision $\varepsilon$ with success probability of $1-\delta$ is:
\begin{equation*}
  64 e^3 \ln(M/\delta) k(2k+2e)^k N^k \varepsilon^{-2}.
\end{equation*}

On the other hand, exact formulas for the second-quantized algorithm~\cite{hugginsNearlyOptimalQuantum2022} shown in the main manuscript are not provided.
We derive the number of ground-state preparation circuits, such that the expectation value of each of the $\mathcal{M}$ observables can be estimated to precision $\varepsilon$ with success probability of $1-\delta$, to be:
\begin{equation}
  (2m+S(2\ln m + 2))\times \frac{\ln(\mathcal{M}/\delta)}{2\left(p-\frac{1}{2}\right)^2}\times \frac{40\ln(10/\varepsilon_2)}{\ln \ln(10/\varepsilon_2)}.\label{eqn:second_quant_measure}
\end{equation}
Furthermore,
\begin{itemize}
  \item $c=2$ and $p=2/3$,
  \item $m=\ln(c\sqrt{\mathcal{M}}/\varepsilon_1)$,
  \item $r^{-1} = 9cm\sqrt{\mathcal{M}}(81\times 8\times 42\pi cm\sqrt{\mathcal{M}}/\varepsilon_1)^{1/2m}$,
  \item $S = 2\pi\times 2^{\lceil \log_2 (4/r\varepsilon_1)\rceil}$,
  \item $\varepsilon_1^2 + \varepsilon_2^2 =\varepsilon$ chosen to minimize the number of ground-state preparation circuits.
\end{itemize}
For the calculation of the 2-RDM, we set $\mathcal{M} = \frac{M(M+1)(M^2+M+2)}{8}$.

\autoref{eqn:second_quant_measure} is derived as follows.
The paper proposing the second-quantized measurement algorithm~\cite{hugginsNearlyOptimalQuantum2022} constructs a probability oracle $U_f$, defined as:
\begin{equation*}
  U_f:\ket{x}\ket{0} \mapsto \ket{x} (\sqrt{f(x)}\ket{1}\ket{\phi_1(x)} + \sqrt{1-f(x)}\ket{0}\ket{\phi_0(x)}),
\end{equation*}
such that the expectation values of the $\mathcal{M}$ observables can be obtained by calculating the gradient of the function $f$:
\begin{equation*}
  \ev*{\hat{O}_i} = \pdv{f}{x_i}.
\end{equation*}
Each application of $U_f$ here requires one application of the ground-state preparation circuit.
\citett{gilyenOptimizingQuantumOptimization2019} was then used to obtain the gradient of $f$.
This work uses a phase oracle $O_f$, defined as:
\begin{equation*}
  O_f:\ket{x}\ket{0}\mapsto e^{if(x)}\ket{x}\ket{0},
\end{equation*}
instead of the probability oracle $U_f$ to obtain a description of the function $f$.
Consequently, the probability oracle needs to be converted to a phase oracle prior to calculating the gradient.
As a result, we divide our derivation here into two steps: obtaining the number of applications of the phase oracle $O_f$ to calculate the gradient, and the number of applications of the probability oracle $U_f$ to calculate the phase oracle.
This explains the different error parameters $\varepsilon_1$ and $\varepsilon_2$ for the two steps, respectively.

First, we obtain the number of applications of the phase oracle to calculate the gradient to precision $\varepsilon_1$.
Following~\citett{gilyenOptimizingQuantumOptimization2019}, we use the notation $O^S_f$ for
\begin{equation*}
  O^S_f:\ket{x}\ket{0}\mapsto e^{iSf(x)}\ket{x}\ket{0}.
\end{equation*}
Looking at Lemma 20 and Theorem 21 of~\citett{gilyenOptimizingQuantumOptimization2019}, the gradient can be calculated using Algorithm 2 of~\citett{gilyenOptimizingQuantumOptimization2019} with success probability of $p=2/3$ with one application of $O^S_f$ with $S=2\pi 2^{\lceil\log_2(4/r\varepsilon_1)\rceil}$.
We then use the median trick to increase the success probability to $1-\delta$ using $\frac{\ln(\mathcal{M}/\delta)}{2(p-1/2)^2}$ repetitions of Algorithm 2, and hence the same number of applications of $O^S_f$.
Next, we move on to Theorem 25 of~\citett{gilyenOptimizingQuantumOptimization2019}, where one application of $O^S_f$ requires $2m+S(2\ln m+2)$ applications of $O_f$, with $m=\ln(c\sqrt{\mathcal{M}}/\varepsilon_1)$.
Theorem 25 also shows shows the exact formula for $r$ to bound the error to $\varepsilon_1$:
\begin{equation*}
  r^{-1} = 9cm\sqrt{\mathcal{M}}(81\times 8\times 42\pi cm\sqrt{M}/\varepsilon_1)^{1/2m}.
\end{equation*}
On the other hand, we have $c=2$ from the definition of the probability oracle~\cite{hugginsNearlyOptimalQuantum2022}.
In total, the gradient requires the following number of applications of $O_f$:
\begin{equation*}
  2m+S(2\ln m+2)\times \frac{\ln(\mathcal{M}/\delta)}{2(p-1/2)^2}
\end{equation*}

Next we follow the procedure described in Section 4.2 of~\citett{gilyenOptimizingQuantumOptimization2019} to obtain the number of applications of the probability oracle $U_f$ required to calculate the phase oracle $O_f$ to precision $\varepsilon_2$.
We assume access to both the ground-state preparation circuit along with its adjoint circuit.
\begin{itemize}
  \item $G_U$ in Equation 10 of~\citett{gilyenOptimizingQuantumOptimization2019} require 2 applications of $U_f$.
  \item $V$ in Equation 17 of~\citett{gilyenOptimizingQuantumOptimization2019} requires $M$ applications of $G_U$ and $M$ applications of $G_U^\dagger$. In this case, we have $M=2\ln(10/\varepsilon_2)/\ln\ln(10/\varepsilon_2)$, shown in the proof of Theorem 14 of~\citett{gilyenOptimizingQuantumOptimization2019}.
  \item $V'=V\otimes R$ in Equation 18 of~\citett{gilyenOptimizingQuantumOptimization2019} requires one application of $V$.
  \item Finally, $O_f$ is obtained using two orders of oblivious amplitude amplification of $V'$ (Theorem 14 of~\citett{gilyenOptimizingQuantumOptimization2019}). This requires 3 applications of $V'$ and 2 applications of $V'^{\dagger}$ as shown in Corollary 42 of~\citett{gilyenOptimizingQuantumOptimization2019}.
\end{itemize}
In total, the phase oracle $O_f$ can be obtained with the following applications of $U_f$ or $U_f^\dagger$:
\begin{equation*}
  \frac{40\ln(10/\varepsilon_2)}{\ln\ln(10/\varepsilon_2)}.
\end{equation*}
Multiplying the two together, we obtain \autoref{eqn:second_quant_measure}.

\section{Comparison between the classical shadows algorithm}
\label{sec:classical_shadows}
In the main manuscript, we chose to highlight the first-quantized classical shadows algorithm of \citett{babbushQuantumSimulationExact2023}.
On the other hand, there is also another classical shadows algorithm which can be performed in the second quantization by \citett{lowClassicalShadowsFermions2024}.
The difference between the two papers lies in their choice of random unitaries and their complexity analysis.
The algorithm proposed by \citett{lowClassicalShadowsFermions2024} uses random basis transformation matrices as the random unitaries to build their classical shadows.
As such basis transformation matrices conserves the occupation number of the wavefunction, the classical shadows only captures the $N$-electron subspace of the whole system, even for $\mathcal{O}(M)$ encodings such as the Jordan-Wigner.
On the other hand, since the first-quantized encoding only encodes the $N$-electron subspace, the classical shadows algorithm of \citett{babbushQuantumSimulationExact2023} can make do with a uniform distribution of the Clifford group over $\lceil \log_2 M\rceil$ qubits, greatly simplifying the construction of such random unitaries.
To implement the classical shadows technique of \citett{lowClassicalShadowsFermions2024} in the first-quantized encoding, we simply replace the uniform distribution of Clifford group with $\lceil \log_2 M\rceil\times\lceil \log_2 M\rceil$ Haar random unitaries.
On the other hand, the same technique for the sorted-list encoding can be implemented by first converting it to the first-quantized encoding via Theorem~\ref{thm:conversion}.

The other difference between the two is in their complexity analysis.
\citett{lowClassicalShadowsFermions2024} estimates the number of ground-state preparation required to bound the average error across all elements of the $k$-RDM.
On the other hand, the number of ground-state preparations estimated by \citett{babbushQuantumSimulationExact2023} ensures that the error of every element of the $k$-RDM is bounded by $\varepsilon$.
This is more compatible with the error guarantees of the gradient-based method~\cite{hugginsNearlyOptimalQuantum2022} we used for the second-quantized measurement algorithm.

\section{Simulation Details for the \texorpdfstring{\ce{Fe2O3}}{Fe2O3} Surface Calculation}
\label{sec:fe2o3_surface_methods}
We performed the DFT calculation of the \ce{Fe2O3} $(01\overline{1}2)$ surface with a benzene molecule, using the plane-wave code Quantum Espresso~\cite{giannozziQUANTUMESPRESSOModular2009,giannozziAdvancedCapabilitiesMaterials2017}.
For the \ce{Fe2O3} $(01\overline{1}2)$ surface, we prepared a \qtyproduct{15.07x16.28x5.65}{\angstrom} system with the norm-conserving pseudopotential and a wavefunction cut-off of \qty{80}{\Ry}.
Furthermore, a DFT+U correction of \qty{5.0}{\eV} was applied to the Fe-$3d$ orbitals.
Geometry optimizations were conducted at varying simulation cell heights ($z$ axis), achieving convergence in the DFT energies for a simulation cell height of \qty{14}{\angstrom}.
We then placed a benzene molecule midway between the surface and its periodic image, and performed geometry optimisations at different simulation cell heights.
In this case, convergence in DFT energies was achieved for a simulation cell height of \qty{23}{\angstrom}, implying an extra \qty{9}{\angstrom} of vacuum space to create unentangled ground states.
Both of these are shown in Figure~4 of the main manuscript.

The \qty{9}{\angstrom} vacuum space corresponds to approximately $\sim$\num{6.5E5} additional plane-waves.
To confirm that the extra vacuum space creates unentangled ground states, we performed an isolated benzene calculation with the same parameters and vacuum space as the isolated \ce{Fe2O3} surface.
The difference in DFT energies between the isolated and the combined systems was 0.899 mRy, well below chemical accuracy.

It is important to note that the plane-wave cut-offs for the many-body quantum simulation are expected to be higher than those for DFT, as the use of pseudopotentials and the approximate exchange-correlation functionals reduce the need for high-frequency wavevectors in DFT.
Furthermore, errors due to the divergence of the coulomb terms are also minimized in DFT as long as the simulation cell is neutral~\cite{martinElectronicStructureBasic2004}.

\section{Details on the Green's function calculation}
\label{sec:greens_methods}
Here, we use the preconditioned fast inversion algorithm~\cite{tongFastInversionPreconditioned2021}.
To estimate the cost for the first-quantized plane-wave Hamiltonian, we follow Theorem 2 of that paper.
We can take the kinetic terms as $U_A$ and the external potential and Coulomb terms as $U_B$, with the 1-norm scaling as $\lambda_A = \mathcal{O}(NM^{2/3}/\Omega^{2/3})$ and $\lambda_B = \mathcal{O}(N^2 M^{1/3}/\Omega^{1/3})$, respectively~\cite{suFaultTolerantQuantumSimulations2021}.
As a result, we can obtain a block-encoding of $(z\pm[H-E_0])^{-1}$ for some $\eta \ge |z|$ using
\begin{align*}
  \tilde{\mathcal{O}}\left( \frac{\lambda_B}{\widetilde{\sigma}_{\text{min}}^2\varepsilon} \ln\left( \frac{1}{\varepsilon \widetilde{\sigma}_{\text{min}}} \right)\ln\left( \frac{1}{\delta} \right)\right)
  &= \tilde{\mathcal{O}}\left( \frac{\lambda_B^3}{\eta^2\varepsilon}\ln\left( \frac{1}{\delta}\right) \right)\\
  &= \tilde{\mathcal{O}}\left(\frac{N^6M}{\Omega \eta^2\epsilon} \ln\left(\frac{1}{\delta}\right)\right)
\end{align*}
applications of $U_A'$ and $U_B$, using the worst-case bound of $\widetilde{\sigma}_{\text{min}} = \eta/(1+\eta+\lambda_B)$.

On the other hand, a block-encoding of $(z\pm[H-E_0])^{-1}$ for some $\eta \ge |z|$ using the non-preconditioned inversion method~\cite{childsQuantumAlgorithmSystems2017} requires
\begin{equation*}
  \tilde{\mathcal{O}}\left( \frac{\lambda}{\eta^2\varepsilon} \right) = \tilde{\mathcal{O}}\left( \frac{NM^{2/3}}{\Omega^{2/3}\eta^2\varepsilon} + \frac{N^2M^{1/3}}{\Omega^{1/3}\eta^{2}\varepsilon} \right)
\end{equation*}
applications of the qubitization circuits, with $\lambda = \lambda_A + \lambda_B$.

\hypertarget{sec:thm_1_extended_proof}{\section*{Supplementary Methods}}
\begin{proof}[Extended Proof for Theorem 1]
  The conversion from the sorted-list encoding to the first-quantized encoding can be performed using the antisymmetrization circuit~\cite{berryImprovedTechniquesPreparing2018}.
  In total, they reported a gate cost of $\mathcal{O}(N\log N\log M)$ and an ancilla cost of $\mathcal{O}(N\log N)$.
  The gate cost here is dominated by the implementation of a reverse sorting network which applies a $Z$ gate when any two registers are swapped.
  It is important to note that the ancilla qubits output by this circuit require postselection.
  In this case, the postselection comes from step 3 of the original implementation~\cite{berryImprovedTechniquesPreparing2018}.
  By this step, a superposition of all length-$N$ strings called \verb|seed| is prepared and sorted.
  However, it still contains repeated entries which must be removed.
  As \verb|seed| is already sorted, this is performed by comparing adjacent entries and we accept when no repetition is detected.
  Hence, this conversion circuit is not unitary within the subspace of the wavefunction qubits, requiring the construction of a different circuit for the reverse conversion.

  On the other hand, the conversion from the first-quantized encoding to the sorted-list encoding is simpler.
  Given an antisymmetrized first-quantized state, we would simply sort the states in an ascending order through the use of a sorting network shown in the following equation
  \begin{equation}
    \sum_{\sigma\in S_N}\frac{\operatorname{sgn}(\sigma)}{\sqrt{N!}}\ket{i_{\sigma(1)}}\otimes\ket{i_{\sigma(2)}}\otimes\cdots\otimes\ket{i_{\sigma(N)}}\otimes\ket{0}\xRightarrow{\text{sort}}\ket{i_1}\otimes\ket{i_2}\otimes\cdots\otimes\ket{i_N}\otimes \left(\sum_{\sigma\in S_N} \frac{1}{\sqrt{N!}}\ket{\text{ancilla}^{\sigma}}\right),\label{eqn:thm1_eqn}
  \end{equation}
  with $i_1 < i_2 < \ldots < i_N$.
  Here, the left hand side represents the first-quantized representation of a Slater determinant with $S_N$ representing the symmetric group of $N$ symbols, and $\operatorname{sgn}(\sigma)$ denotes the sign of the permutation.
  Similarly with the antisymmetrization circuit, a $Z$ gate is also applied when any two registers are swapped to account for the $\operatorname{sgn}(\sigma)$ term.
  The sorted-list state shown on the right-hand side above is obtained after the registers are sorted.
  As a result, the conversion from the first-quantized encoding to the sorted-list encoding also costs $\mathcal{O}(N\log N\log M)$ gates and $\mathcal{O}(N\log N)$ ancillae.

  Although no postselection of ancilla qubits is required in this case, the ancilla qubits used in this conversion circuit is also not necessarily in the $\ket{0}$ state upon output due to the sorting network.
  These ancilla qubits originate from the comparators used in the sorting network (see \autoref{fig:comparison}), which outputs a $\ket{1}$ when the two input registers are swapped and $\ket{0}$ otherwise.
  Crucially, these output ancilla qubits are unentangled from the wavefunction qubits, provided the input is a valid first-quantized wavefunction.
  Consider a single Slater determinant as an input as shown in \autoref{eqn:thm1_eqn}.
  As each state $\ket{i_{\sigma(1)}}\otimes\ket{i_{\sigma(2)}}\otimes\cdots\otimes\ket{i_{\sigma(N)}}$ is all sorted back to $\ket{i_1}\otimes\ket{i_2}\otimes\cdots\otimes\ket{i_N}$, it is disentangled from the ancilla qubits.
  The same holds true when the input is a linear superposition of Slater determinants:
  \begin{gather}
    \sum_l \sum_{\sigma\in S_N}\frac{\alpha_l\operatorname{sgn}(\sigma)}{\sqrt{N!}}\ket{i_{\sigma(1)}^{(l)}}\otimes\ket{i_{\sigma(2)}^{(l)}}\otimes\cdots\otimes\ket{i_{\sigma(N)}^{(l)}}\otimes \ket{0}\xRightarrow{\text{sort}}\nonumber\\
    \sum_l \alpha_l \ket{i_1^{(l)}}\otimes \ket{i_2^{(l)}}\otimes\cdots\otimes \ket{i_N^{(l)}}\otimes \left( \sum_{\sigma\in S_N}\frac{1}{\sqrt{N!}} \ket{\text{ancilla}^{\sigma}} \right).\label{eqn:thm1_eqn2}
  \end{gather}
  This is because the swapping performed by the comparators depends solely on the relative orders of the orbital indices.
  Since $i^{(l)}_1<i^{(l)}_2<\ldots<i^{(l)}_N$ for all $l$, such specific sequence of swaps, and thus the state of $\ket{\text{ancilla}^{\sigma}}$, depends only on the permutation index $\sigma$ as shown in \autoref{eqn:thm1_eqn} and~\ref{eqn:thm1_eqn2}.
  Consequently, the output state of the ancilla qubits, $\sum_{\sigma\in S_N} \ket{\text{ancilla}^{\sigma}}$, is a fixed, unentangled state that depends only on $N$, even when the input is a linear combination of Slater determinants.

  \begin{figure}[h]
    \centering
    \includegraphics{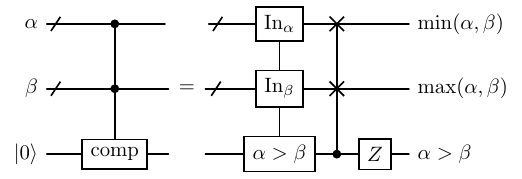}
    \caption{
      Comparators used as a building block of sorting network circuits.
      This performs a swap on registers $\alpha$ and $\beta$ if $\alpha > \beta$.
      The ancilla qubits outputs a $\ket{1}$ state if the swap is performed and $\ket{0}$ otherwise.
      A $Z$ gate is also added to account for parity when implemented for the conversion circuit of Theorem~\ref{thm:conversion}.
    }
    \label{fig:comparison}
  \end{figure}

  \begin{figure}[h]
    \centering
    \includegraphics{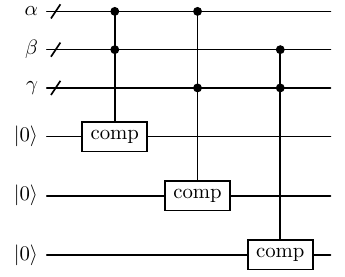}
    \caption{
      Circuit used to convert from the first quantization to the second quantization for $N=3$.
    }
    \label{fig:comparison_3elec}
  \end{figure}

  As an illustrative example, consider the $N=3$ case, with the circuit shown in \autoref{fig:comparison_3elec}.
  First, consider a single Slater determinant as the input:
  \begin{equation*}
    \ket{\Psi_{\text{fq}}} = \frac{1}{\sqrt{6}}[\ket{a,b,c} - \ket{a,c,b} - \ket{b,a,c} + \ket{b,c,a} - \ket{c,b,a} + \ket{c,a,b}],
  \end{equation*}
  with $a < b < c$.
  After applying the circuit of \autoref{fig:comparison_3elec}, all the permutations of $a,b,c$ will be sorted back to the ascending order, yielding a valid sorted-list representation disentangled from the ancilla qubits:
  \begin{equation*}
    \ket{a,b,c}\otimes \frac{1}{\sqrt{6}}[\ket{000}+\ket{010}+\ket{100}+\ket{011}+\ket{111}+\ket{101}].
  \end{equation*}
  As the ancilla qubits above does not depend on $a$, $b$, and $c$, the disentangled property holds true when the input is a linear superposition of Slater determinants
  \begin{align*}
    \ket{\Psi_{\text{fq}}} &= \sum_i\frac{\alpha_i}{\sqrt{6}}[\ket{a_i,b_i,c_i} - \ket{a_i,c_i,b_i} - \ket{b_i,a_i,c_i} + \ket{b_i,c_i,a_i} - \ket{c_i,b_i,a_i} + \ket{c_i,a_i,b_i}]\\
                           &\Rightarrow \left(\sum_i\alpha_i \ket{a_i,b_i,c_i}\right)\otimes\frac{1}{\sqrt{6}}[\ket{000}+\ket{010}+\ket{100}+\ket{011}+\ket{111}+\ket{101}].\qedhere
  \end{align*}
\end{proof}
\end{document}